\documentclass[11pt]{article}
\usepackage{axodraw}
\usepackage{epsfig}
\usepackage{amsfonts}
\usepackage{amsmath}
\usepackage{bbm}
\input pix.sty

\renewcommand{\eq}{eq.~}
\renewcommand{\eqs}{eqs.~}
\renewcommand{\se}{sec.~}

\renewcommand{\fig}{fig.~}

\newcommand{\tinymsbar}{{\overline{\mbox{\tiny\rm{MS}}}}}
\newcommand{\Lambdamsbar}{{\Lambda_\tinymsbar}}

\newcommand{\Nf}{N_{\rm f}}
\newcommand{\Nc}{N_{\rm c}}

\newcommand{\gB}{g_\rmii{B}}

\newcommand{\gammaE}{\gamma_\rmii{E}}

\newcommand{\rmO}{{\mathcal{O}}}
\newcommand{\bmu}{\bar\mu}

\def\lsi{\raise0.3ex\hbox{$<$\kern-0.75em\raise-1.1ex\hbox{$\sim$}}}
\def\gsi{\raise0.3ex\hbox{$>$\kern-0.75em\raise-1.1ex\hbox{$\sim$}}}
\newcommand{\lsim}{\mathop{\lsi}}
\newcommand{\gsim}{\mathop{\gsi}}

\newcommand{\sign}{\mathop{\mbox{sign}}}
\newcommand{\nF}{n_\rmii{F}}
\newcommand{\nB}{n_\rmii{B}}
 \renewcommand{\nF}[1]{n_\rmii{F{#1}}}
 \renewcommand{\nB}[1]{n_\rmii{B{#1}}}
\newcommand{\rmii}[1]{{\mbox{\tiny\rm{#1}}}}
\newcommand{\re}{\mathop{\mbox{Re}}}
\newcommand{\im}{\mathop{\mbox{Im}}}

\newcommand{\Tint}[1]{{\hbox{$\sum$}\!\!\!\!\!\!\!\int\,}_{\!\!\!\!\raise-0.9ex\hbox{$\scriptstyle{#1}$}}}
\newcommand{\Tinti}[1]{{{\Sigma}\!\!\!\!\raise0.3ex\hbox{$\int$}_\rmii{${#1}$}}}

\newcommand{\unit}{{\mathbbm{1}}} %{\ii}
\newcommand{\bi}{\begin{itemize}}
\newcommand{\ei}{\end{itemize}}
\newcommand{\hide}[1]{ }

\def\TAsc(#1,#2)(#3,#4,#5)%
{\SetWidth{2.0}\CArc(#1,#2)(#3,#4,#5)\SetWidth{1.0}}
\def\Lwidth{3}

\def\TAgl(#1,#2)(#3,#4,#5){\SetWidth{2.0}\PhotonArc(#1,#2)(#3,#4,#5){\Lwidth}%
{6.283 #3 mul 360 div #4 #5 sub #4 #5 sub mul sqrt mul Tdensity mul}%
\SetWidth{1.0}}
\def\TLgl(#1,#2)(#3,#4){\SetWidth{2.0}\Photon(#1,#2)(#3,#4){\Lwidth}
{#1 #3 sub #1 #3 sub mul #2 #4 sub #2 #4 sub mul add sqrt Tdensity mul}%
\SetWidth{1.0}}
\renewcommand{\picc}[1]{\;\parbox[c]{60pt}{\begin{picture}(60,30)(0,0)
\SetWidth{1.0}\SetScale{1.3} #1 \end{picture}}\; }
\def\Lwidth{1.3}

\newcommand{\picu}[1]{\;\parbox[c]{60pt}{\begin{picture}(60,30)(0,0)
\SetWidth{1.0}\SetScale{1.0} #1 \end{picture}}\; }
\def\EleA{\picu{%
 \Agl(30,5)(22.3,27,153)%
 \Agl(30,25)(22.3,207,333)%
 \COval(10,15)(2,2)(0){Black}{Black}%
 \COval(50,15)(2,2)(0){Black}{Black}%
}}
\def\EleB{\picu{%
 \Agl(30,5)(22.3,27,153)%
 \Agl(30,25)(22.3,207,333)%
 \COval(10,15)(2,2)(0){Black}{Black}%
 \COval(50,15)(2,2)(0){Black}{Black}%
 \Agl(58,15)(8,0,360)%
}}
\def\EleC{\picu{%
 \Agl(30,5)(22.3,27,153)%
 \Agl(30,25)(22.3,207,333)%
 \COval(10,15)(2,2)(0){Black}{Black}%
 \COval(50,15)(2,2)(0){Black}{Black}%
 \Lgl(10,15)(50,15)%
}}
\def\EleD{\picu{%
 \Agl(30,5)(22.3,90,153)%
 \Agl(30,25)(22.3,207,333)%
 \COval(10,15)(2,2)(0){Black}{Black}%
 \COval(50,15)(2,2)(0){Black}{Black}%
 \Agl(43,27)(12,180,300)%
 \Agl(38,16)(12,0,120)%
}}
\def\EleE{\picu{%
 \Agl(30,5)(22.3,27,153)%
 \Agl(30,25)(22.3,207,333)%
 \COval(10,15)(2,2)(0){Black}{Black}%
 \COval(50,15)(2,2)(0){Black}{Black}%
 \GCirc(30,27.3){4}{0.5}
}}
\def\EleF{\picu{%
 \Agl(20,10)(11.15,27,153)%
 \Agl(20,20)(11.15,207,333)%
 \Agl(40,10)(11.15,27,153)%
 \Agl(40,20)(11.15,207,333)%
 \COval(10,15)(2,2)(0){Black}{Black}%
 \COval(50,15)(2,2)(0){Black}{Black}%
}}
\def\EleG{\picu{%
 \Agl(30,5)(22.3,27,153)%
 \Agl(30,25)(22.3,207,333)%
 \COval(10,15)(2,2)(0){Black}{Black}%
 \COval(50,15)(2,2)(0){Black}{Black}%
 \Lgl(30,2.7)(30,27.3)%
}}
\makeatletter \@addtoreset{equation}{section} \makeatother
\renewcommand{\theequation}{\arabic{section}.\arabic{equation}}
\makeatletter
\renewcommand\section{\@startsection {section}{1}{\z@}%
                                   {-5.5ex \@plus -1ex \@minus -.2ex}% bfr-
                                   {2.3ex \@plus.2ex}%
                                   {\normalfont\large\bfseries}}
\renewcommand\subsection{\@startsection{subsection}{2}{\z@}%
                                     {-3.25ex\@plus -1ex \@minus -.2ex}%
                                     {1.5ex \@plus .2ex}%
                                     {\normalfont\normalsize\bfseries}}
\renewcommand\thesection {\@arabic\c@section}
\renewcommand\thesubsection   {\thesection.\@arabic\c@subsection}
\renewcommand{\@seccntformat}[1]{%
\csname the#1\endcsname.\hspace{1.0em}}
\makeatother
\begin{document}

\flushbottom

\begin{titlepage}

\begin{flushright}
% OUTLINE  \\ 
% DRAFT \\ 
BI-TP 2010/23\\
% arXiv:1008.3263\\ 
\vspace*{1cm}
\end{flushright}
\begin{centering}
\vfill

{\Large{\bf
 Ultraviolet asymptotics of scalar and pseudoscalar \\[2mm]
 correlators in hot Yang-Mills theory
}} 

\vspace{0.8cm}

M.~Laine$^{\rm a}$, %%\footnote{laine@physik.uni-bielefeld.de}
M.~Veps\"al\"ainen$^{\rm b}$, %%\footnote{mtvepsal@pcu.helsinki.fi}
A.~Vuorinen$^{\rm a}$ %%\footnote{vuorinen@physik.uni-bielefeld.de}

\vspace{0.8cm}

$^\rmi{a}$%
{\em
Faculty of Physics, University of Bielefeld, 
D-33501 Bielefeld, Germany\\}

\vspace{0.3cm}

$^{\rm b}$%
{\em 
Department of Physics, 
P.O.Box 64, FI-00014 University of Helsinki, Finland\\}

\vspace*{0.8cm}

\mbox{\bf Abstract}
 
\end{centering}

\vspace*{0.3cm}
 
\noindent
Inspired by recent lattice measurements, we determine 
the short-distance ($a \ll r \ll 1/\pi T$) as well as large-frequency 
($1/a \gg \omega \gg \pi T$) asymptotics of scalar (trace anomaly) 
and pseudoscalar (topological charge density) correlators at 2-loop
order in hot Yang-Mills theory. The results are expressed in the 
form of an Operator Product Expansion. We confirm and refine 
the determination of a number of Wilson coefficients; however 
some discrepancies with recent literature are detected as well, 
and employing the correct values might help, on the qualitative 
level, to understand some of the features observed 
in the lattice measurements. On the other hand, 
the Wilson coefficients show slow convergence and it appears uncertain 
whether this approach can lead to quantitative comparisons with lattice 
data. Nevertheless, as we outline, our general results might 
serve as theoretical starting points for a number of perhaps
phenomenologically more successful lines of investigation. 

\vfill

%% %\noindent
%% %PACS numbers: 
%% %11.10.Wx, %        Finite temperature field theory
%% { %11.15.Ha, %        Lattice gauge theory } 
%% %12.38.Bx, %        Perturbative calculations in QCD
%% %12.38.Mh, %        Quark--gluon plasma
%% %14.40.Nd, %        Bottom mesons
%% %\\
%% %Keywords:
 
\vspace*{1cm}
  
\noindent
September 2010

\vfill

\end{titlepage}

%%%%%%%%%%%%%%%%%%%%%%%%%%% SECTION %%%%%%%%%%%%%%%%%%%%%%%%%%%%%%%%%%%%%%
%
\section{Introduction}

Many of the most interesting physical properties of 
a finite-temperature system are of an ``infrared'' type, i.e.\ to 
be extracted from the long-distance or short-frequency limit 
of appropriate 2-point correlation functions. For instance, in 
a weakly coupled Yang-Mills theory with the gauge coupling 
$g$, spatial correlation lengths originate 
at length scales $r\gsim 1/(gT)$ or $r\gsim \pi/(g^2T)$, depending
on the global quantum numbers of the operator considered~\cite{ay}.
At the same time, transport coefficients, which reflect 
the real-time response of the system to small 
perturbations, arise at frequencies $\omega \lsim g^4 T/\pi^3$
(cf.,\ e.g.,\ ref.~\cite{ms}). Infrared observables may 
either be genuinely non-perturbative~\cite{linde,gpy}, or they do possess
a weak-coupling series up to some order, but it  
is slowly convergent at temperatures relevant for heavy ion
collision experiments
(cf.,\ e.g.,\ ref.~\cite{chm}). 
Hence, perhaps with a few exceptions
(cf.,\ e.g.,\ ref.~\cite{lv}), 
infrared observables need eventually to 
be determined via non-perturbative lattice simulations. 
 
Despite the stated general picture, circumstances exist 
as well under which ``ultraviolet'' observables, measured at  
short distances ($r \lsim 1/\pi T$) or large frequencies
($\omega \gsim \pi T$), are of physical 
interest. As an example on the former case, we may mention that in 
heavy quarkonium physics, the system has an additional ``external''
scale, the heavy quark mass, $M$. Given that normally $\pi T \ll M$, 
the inverse Bohr radius $r_\rmii{B}^{-1} \sim \alpha_s M \ll M$ 
of a quarkonium state could well be 
of the same order as the temperature, $r_\rmii{B}^{-1} \sim \pi T$.
Changes in quarkonium properties caused by a finite temperature
could therefore be due to thermal modifications of 
the quark--antiquark potential at $r \sim r_\rmii{B} \sim 1/(\pi T)$, 
in which case the weak-coupling expansion may converge 
faster than at large distances $r\gsim 1/(gT)$~\cite{singlet}.
As an example on the latter case, we remark that all lattice
estimates of a spectral function, $\rho(\omega)$, from whose 
intercept, $\lim_{\omega\to 0} \rho(\omega)/\omega$, transport
coefficients are determined, rely on ``inverting'' the relation
\be
 G(\hat \tau) = 
 \int_0^\infty
 \frac{{\rm d}\omega}{\pi} \rho(\omega)
 \frac{\cosh \left(\frac{1}{2} - \hat\tau\right)\beta\omega}
 {\sinh\frac{\beta \omega}{2}} 
 \;, \la{int_rel} 
\ee
where $\beta \equiv 1/T$ and $G(\hat\tau)$, $0< \hat\tau< 1$,  
is a measured Euclidean correlator along the compact time direction. 
Since the ``kernel'' multiplying $\rho(\omega)$ 
in \eq\nr{int_rel} only depends on $\omega$ through $\beta\omega$,
it is clear that $\rho(\omega)$ in
the regime $\omega\sim \pi T$ gives an important 
contribution to $G(\hat\tau)$, and needs to be well understood before 
infrared sensitive contributions from the range $\omega \ll \pi T$ can 
be reliably extracted. 

In the present note we study the ultraviolet regime
of certain 2-point correlation functions, 
and even take it to its
extreme limit: not only do we consider $r\lsim 1/(\pi T)$ but 
in fact $r \ll 1/(\pi T)$; 
not only $\omega\gsim \pi T$ but in fact $\omega \gg \pi T$.
Only the ultraviolet cutoff, e.g.\ the lattice
spacing $a$, is assumed to be even farther in the ultraviolet than 
our physical scales. In this situation the correlation functions can 
be determined within a framework similar to 
the Operator Product Expansion~\cite{kgw}, 
as has recently been discussed in the Euclidean 
domain~\cite{hbm_b} and also more generally~\cite{sch}.
Apart from theoretical considerations, 
we would in principle also like to make contact 
with the Euclidean lattice simulations described 
in ref.~\cite{hbm_c}. It is our experience, both from 
Euclidean~\cite{singlet} and, after analytic continuation, 
Minkowskian~\cite{rhoE} domains that, 
when pursued to a sufficient order, the (continuum)
weak-coupling expansion could work relatively 
well in the ultraviolet regime; however, the issue needs
to be re-investigated when a new correlation function 
is considered or when the computation is organized
as an Operator Product Expansion, and these
are some of the goals of the present study. 

The plan of this note is the following. 
In \se\ref{se:setup} we define the basic observables considered. 
Section~\ref{se:method} contains an outline of the method used; 
the general results are given in \se\ref{se:mom}.
The case of short distances is discussed more specifically
in \se\ref{se:r}, and that of large frequencies in \se\ref{se:w}. 
We also briefly comment on the relation of our work to 
recently discussed ``sum rules'' in \se\ref{se:sum}. 
Section~\ref{se:concl} offers a summary and outlook, whereas 
the three appendices collect together some details needed in the main text: 
the definitions and asymptotic expansions of the ``master''
sum-integrals appearing in pure Yang-Mills theory are listed in appendix A; 
a number of thermodynamic potentials playing a role in our study 
are given in appendix B; and fermionic effects are briefly 
discussed in appendix C.

%%%%%%%%%%%%%%%%%%%%%%%%%%%%% SECTION %%%%%%%%%%%%%%%%%%%%%%%%%%%%%%%%%%%%
%
\section{Setup}
\la{se:setup}

Employing the convention
$
 D_\mu = \partial_\mu - i \gB A^a_\mu T^a
$, 
with $T^a$ hermitean generators of SU($\Nc$) normalized as 
$
 \tr[T^a T^b] = \fr12 \delta^{ab}
$, 
and defining 
$
 F^a_{\mu\nu} = (2i/\gB) \tr \{ T^a [D_\mu, D_\nu] \}
 = 
 \partial_\mu A^a_\nu - \partial_\nu A^a_\mu + \gB f^{abc} A^b_\mu A^c_\nu 
$, 
the dimensionally regularized 
Euclidean action relevant for pure Yang-Mills theory
at a finite temperature $T = 1/\beta$ reads
\be
 S_E = \int_{0}^{\beta} \! {\rm d}\tau \int \! {\rm d}^{3-2\epsilon}\vec{x}
 \, \left\{ \frac{1}{4} F^a_{\mu\nu} F^a_{\mu\nu} \right\}
 \;. \la{SE}
\ee
The trace of the corresponding energy-momentum tensor is 
$\frac{D-4}{4} F^a_{\mu\nu} F^a_{\mu\nu}$, where $D\equiv 4-2\epsilon$ is 
the space-time dimensionality. In order to define the operators whose 
correlation functions we are interested in, 
we note that it is the ``geometric'' structure 
$
 \gB^2 F^a_{\mu\nu}F^a_{\rho\sigma} = 
 -2 \tr \{ [D_\mu, D_\nu] [D_\rho, D_\sigma] \}
$
which requires no renormalization at the order of our 
computation. So, we define the gauge invariant
scalar and pseudoscalar operators 
\be
 \theta \equiv c_\theta\, \gB^2 F^a_{\mu\nu}F^a_{\mu\nu}
 \;, \quad
 \chi \equiv c_\chi\, \epsilon_{\mu\nu\rho\sigma} 
 \gB^2 F^a_{\mu\nu}F^a_{\rho\sigma}
 \;, \la{ops}
\ee
where the $D$-dimensional Euclidean $\epsilon_{\mu\nu\rho\sigma}$ is handled
as in ref.~\cite{old} (as long as the procedure is 
self-consistent the precise regularization
has no impact on the final results). 
In analogy with ref.~\cite{hbm_c}
the coefficients are chosen as
\ba
 c_\theta & = & \frac{D-4}{4\gB^2\mu^{-2\epsilon}} = 
 -\frac{b_0}{2} - \frac{b_1 g^2}{4} + \ldots 
 \;, \la{c_theta} \\ 
 c_\chi & \equiv & \frac{1}{64\pi^2} 
 \;, \la{c_chi}
\ea
but in most of what follows we do not need to specify their values.
For \eq\nr{c_theta}, we have written 
\be
 \gB^2 = g^2 \mu^{2\epsilon} \biggl[
 1 - \frac{b_0 g^2}{\epsilon} + 
 \biggl( \frac{b_0^2}{\epsilon^2} - \frac{b_1}{2\epsilon} \biggr) g^4 + \ldots 
 \biggr] 
 \;, \la{gB}
\ee
where
\be
 b_0 = \frac{11\Nc}{3(4\pi)^2}
 \;, \quad 
 b_1 = \frac{34\Nc^2}{3(4\pi)^4}
 \;, \la{b0}
\ee
and $g^2$ %% \equiv g^2(\bmu) \mu^{-2\epsilon}$ 
is the dimensionless
renormalized coupling constant, 
evaluated at the $\msbar$ scheme renormalization scale $\bmu$
($ 
 \mu^{2} =  \bmu^{2}
 {e^{\gammaE}}/{4\pi}
$).
% As usual we denote $\alpha_s \equiv g^2/4\pi$. 

With this notation, the Euclidean correlators considered are defined as 
\be
 G_\theta(x) \equiv  
 \langle \theta(x) \theta(0) \rangle_\rmi{c}
 \;, \quad
 G_\chi(x) \equiv  
 \langle \chi(x) \chi(0) \rangle
 \;, \la{Gx}
\ee
where $\langle ... \rangle_\rmi{c}$ denotes the connected part, 
and the expectation value is taken at a finite temperature $T$. 
(The disconnected part of $G_\theta$
is 
$\langle \theta \rangle^2
$, 
where $\langle \theta \rangle = e - 3 p$ 
is trace of the energy-momentum tensor; in a general regularization
scheme $\langle \theta \rangle$ is 
ultraviolet divergent. In the following we need the finite thermal
part thereof, denoted by $(e-3p)(T)$.) 
We also consider the corresponding Fourier transforms, 
\be
 \tilde G_\theta(P) \equiv \int_x e^{- i P\cdot x} G_\theta(x) 
 \;, \quad
 \tilde G_\chi(P) \equiv \int_x e^{- i P\cdot x} G_\chi(x) 
 \;, \la{GP}
\ee
where
short-distance singularities are regulated dimensionally.\footnote{%
 For a careful general analysis of short-distance 
 singularities in the case of $\tilde G_\chi$, see ref.~\cite{ml}.
 } 
Finally we denote 
\be
 \Delta \tilde G_\theta(P) \equiv \tilde G_\theta(P) - \tilde G_\theta^{T=0}(P)
 \;, \quad
 \Delta \tilde G_\chi(P) \equiv \tilde G_\chi(P) - \tilde G_\chi^{T=0}(P)
 \;, 
\ee
subtracting the zero-temperature parts at a fixed $P$. 
In Euclidean signature, 
the components of the four-momentum and spacetime coordinate
are denoted by 
\be
 P = (p_n,\vec{p})\;, \quad
 p \equiv |\vec{p}|\;, \quad
 x = (\tau,\vec{x})\;, \quad
 r \equiv |\vec{x}|\;, \la{def_P_x}
\ee 
where $p_n = 2\pi T n$ are bosonic Matsubara frequencies.

%%%%%%%%%%%%%%%%%%%%%%%%%%%% SECTION %%%%%%%%%%%%%%%%%%%%%%%%%%%%%%%%%%%%
%
\section{Method}
\la{se:method}

Carrying out the Wick contractions in the 1-loop and 2-loop 
Feynman graphs contributing to \eq\nr{Gx}
(cf.\ ref.~\cite{old} or \fig\ref{fig:graphs}) and using the notation
of appendix~A for the various ``master'' sum-integrals
appearing in the result, 
we find the bare expressions ($d_A \equiv \Nc^2-1$) 
\ba
 && \hspace*{-1cm} \frac{\tilde G_\theta(P)}{4 d_A c_\theta^2 \gB^4} = 
% \nn & = & 
 (D-2) \biggl[ -\mathcal{J}_\rmi{a} + \fr12 \mathcal{J}_\rmi{b} \biggr]
 \nn 
 & + & \gB^2 \Nc \biggl\{ 
 2 (D-2) \biggl[ - (D-1)\mathcal{I}_\rmi{a} + (D-4) \mathcal{I}_\rmi{b} \biggr]
 + (D-2)^2 \biggl[ \mathcal{I}_\rmi{c} - \mathcal{I}_\rmi{d} \biggr]
 \nn & & \quad + \, 
 \frac{22-7D}{3} \mathcal{I}_\rmi{f} - \frac{(D-4)^2}{2} \mathcal{I}_\rmi{g}
 + (D-2) 
 \biggl[ 
   -3 \mathcal{I}_\rmi{e} + 3 \mathcal{I}_\rmi{h} + 2\mathcal{I}_\rmi{i}
    - \mathcal{I}_\rmi{j}
 \biggr] \biggr\} % + \rmO(\gB^4)
 \;, \la{Gtheta_bare} \\
%%%%%%%%%%%%%%%%%%%%%%%%%%%%%%%%%%%%%%
%%%%%%%%%%%%%%%%%%%%%%%%%%%%%%%%%%%%%% 
 && \hspace*{-1cm} \frac{\tilde G_\chi(P)}{-16 d_A c_\chi^2 \gB^4 (D-3)} = 
% \nn & = & 
 (D-2) \biggl[ -\mathcal{J}_\rmi{a} + \fr12 \mathcal{J}_\rmi{b} \biggr]
 \nn 
 & + & \gB^2 \Nc \biggl\{ 
 2 (D-2) \biggl[ - \mathcal{I}_\rmi{a} + (D-4) \mathcal{I}_\rmi{b} \biggr]
 + (D-2)^2 \biggl[ \mathcal{I}_\rmi{c} - \mathcal{I}_\rmi{d} \biggr]
 \nn & & \quad - \, 
 \frac{2 D^2-17D+42}{3} \mathcal{I}_\rmi{f} - 2 (D-4) \mathcal{I}_\rmi{g}
 + (D-2) 
 \biggl[ 
   -3 \mathcal{I}_\rmi{e} + 3 \mathcal{I}_\rmi{h} + 2\mathcal{I}_\rmi{i}
    - \mathcal{I}_\rmi{j}
 \biggr] \biggr\} % + \rmO(\gB^4)
 \;. \la{Gchi_bare}
\ea 
Here, for brevity, structures containing $\Tinti{Q} 1$, which 
vanishes exactly in dimensional regularization, have been omitted 
($\Tinti{Q}$ denotes a sum-integral with bosonic Matsubara frequencies).

%%%%%%%%%%%%%%%%%%%%%%%%% FIGURE %%%%%%%%%%%%%%%%%%%%%%%%%%%%%%%%%%%%%%%%%
%
\begin{figure}[t]
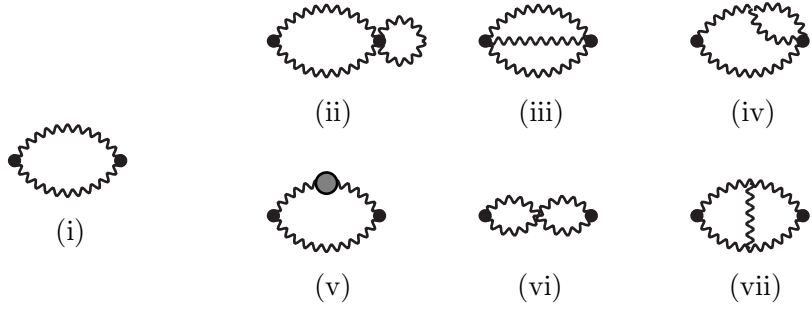


\hspace*{1.5cm}%
\begin{minipage}[c]{3cm}
\begin{eqnarray*}
&& 
 \hspace*{-1cm}
 \EleA 
\\[1mm] 
&& 
 \hspace*{0.0cm}
 \mbox{(i)} 
\end{eqnarray*}
\end{minipage}%
\begin{minipage}[c]{10cm}
\begin{eqnarray*}
&& 
 \hspace*{-1cm}
 \EleB \quad\; 
 \EleC \quad\; 
 \EleD \quad\; 
\\[1mm] 
&& 
 \hspace*{0.0cm}
 \mbox{(ii)} \hspace*{2.2cm}
 \mbox{(iii)} \hspace*{2.2cm}
 \mbox{(iv)} 
\\[5mm] 
&& 
 \hspace*{-1cm}
 \EleE \quad\; 
 \EleF \quad\; 
 \EleG \quad 
\\[1mm] 
&& 
 \hspace*{0.0cm}
 \mbox{(v)} \hspace*{2.2cm}
 \mbox{(vi)} \hspace*{2.2cm}
 \mbox{(vii)} 
 \\[3mm]
\end{eqnarray*}
\end{minipage}

\caption[a]{\small 
The graphs contributing to the connected 2-point correlators 
defined in \eq\nr{Gx}, up to 2-loop order. 
The wiggly lines denote gluons; the small dots the 
operators $\theta$ or $\chi$ (cf.\ \eq\nr{ops}); 
and the grey blob the 1-loop gauge field self-energy. Graphs obtained 
with trivial ``reflections'' from those shown have been 
omitted from the figure.} 
\la{fig:graphs}
\end{figure}
%
%%%%%%%%%%%%%%%%%%%%%%%%%%%%%%%%%%%%%%%%%%%%%%%%%%%%%%%%%%%%%%%%%%%%%%%%%%

The goal now is to obtain asymptotic expansions for the functions
in \eqs\nr{Gtheta_bare}, \nr{Gchi_bare}, 
valid in the Euclidean domain $P^2 = p_n^2 + {p}^2 \gg (\pi T)^2$.
(Interestingly, as has recently been reviewed in ref.~\cite{jan}, 
similar expansions may play a role also in the computation 
of high-order vacuum graphs at finite temperature.)

In order to obtain the expansions, we first carry out all 
the Matsubara sums, which can be done exactly. 
This has the effect of setting one or
two of the propagators ``on-shell''; the corresponding line is 
weighed by the Bose distribution. The coefficient can be 
identified as a zero-temperature amplitude or 1-loop integral, 
with special kinematics.\footnote{%
 Various recipes for this can be found in the literature, 
 but we have established all relations from scratch.
 } 
For instance, introducing the notation
\be
 {[ \ldots ]}_Q \equiv 
 \fr12 \sum_{q_n = \pm i q} \{ \ldots \}
 \;, \quad 
 {[ \ldots ]}_{Q,R} \equiv 
 \fr14 \sum_{q_n = \pm i q} \sum_{r_n=\pm i r} \{ \ldots \}
 \;, 
\ee  
it is straightforward to verify that 
the sum-integral $\mathcal{I}_\rmi{h}$ 
(cf.\ \eq\nr{def_Ih})
can be re-expressed as
\ba
 \mathcal{I}_\rmi{h} \!\! & = & \!\! 
 \int_{Q,R} \frac{P^4}{Q^2R^2(Q-R)^2(R-P)^2} 
 \nn & + & 
 \int_{\vec{q}} \frac{\nB{}(q)}{q}
 \nn & & \!\! \times  
 \int_R \biggl[ 
   \frac{2 P^4}{R^2(Q-R)^2(R-P)^2 }
% \nn & & 
% \hphantom{ \int_{\vec{q}} \frac{\nB{}(q)}{q}}
  + \frac{P^4}{R^2(Q-R)^2(Q-P)^2}
% \nn & & 
% \hphantom{ \int_{\vec{q}} \frac{\nB{}(q)}{q}}
  + \frac{P^4 }{(Q-P)^2(Q-R)^2(R-P)^2}
 \biggr]_Q
  \nn & + & 
 \int_{\vec{q},\vec{r}} \frac{\nB{}(q)}{q} \frac{\nB{}(r)}{r}
  \nn & & \!\! \times 
 \biggl[ 
   \frac{2 P^4}{(Q-R)^2(R-P)^2 }
% \nn & & 
% \hphantom{ \int_{\vec{q},\vec{r}} \frac{\nB{}(q)}{q} \frac{\nB{}(r)}{r} }
  +
   \frac{P^4}{(Q-R)^2(Q-R-P)^2 }
% \nn & & 
% \hphantom{ \int_{\vec{q},\vec{r}} \frac{\nB{}(q)}{q} \frac{\nB{}(r)}{r} }
  +
   \frac{2 P^4}{(R-P)^2(Q+R-P)^2 } 
 \biggr]_{Q,R}
 \;, \la{dx_exact} 
\ea
where 
$
 \int_Q \equiv \int {\rm d}^D Q/(2\pi)^D
$ 
and 
$
 \int_{\vec{q}}  \equiv \int {\rm d}^{D-1} \vec{q}/(2\pi)^{D-1}
$.
It is important to realize that within the square brackets we can  
set $Q^2 = 0$ for any $D$, and that therefore anything 
proportional to $D$-dependent powers of $Q^2$ vanishes 
exactly (this is the case
particularly when the $R$-integration factorizes from the $P$-dependence).

In order to handle the remaining structures, we note that 
an integration variable appearing inside the Bose distribution
is always ultraviolet safe (cut off by the temperature). Therefore, 
we can expand propagators as 
\ba
 \biggl[ \frac{1}{(Q-R)^2} \biggr]_Q & = &
 \biggl[ \frac{1}{R^2} + \frac{2Q\cdot R}{R^4} + 
         \frac{4(Q\cdot R)^2}{R^6} + \ldots \biggr]_Q
 \;, 
\ea
where we also made use of the on-shell condition $Q^2=0$. Such an 
expansion leads to the following types of vacuum integrals: 
\ba
 \int_R \frac{R_\mu R_\nu}{(R^2)^m[(R-P)^2]^n} & = & 
 \delta_{\mu\nu} \, A_{m,n}  + P_\mu P_\nu \, B_{m,n} 
 \;, \\
 \int_R \frac{R_\mu }{(R^2)^m[(R-P)^2]^n} & = & 
 P_\mu \, C_{m,n}  
 \;, \\ 
 \int_R \frac{1 }{(R^2)^m[(R-P)^2]^n} & = & 
 I_{m,n} 
 \;. 
\ea
In dimensional regularization these can all be related to $I_{m,n}$, 
which in turn reads
\be
 I_{m,n} = \frac{(P^2)^{\frac{D}{2} - m - n}}{(4\pi)^{\frac{D}{2}}}
 \frac{\Gamma(m+n-\frac{D}{2})\Gamma(\frac{D}{2}-m)\Gamma(\frac{D}{2}-n)}
 {\Gamma(D-m-n)\Gamma(m)\Gamma(n)}
 \;.
\ee
The remaining $P$-dependence often appears in the form 
$
 {[(P\cdot Q)^2]}_Q = q^2 (\frac{p^2}{3-2\epsilon} - {\scriptstyle p_n^2})
$, 
where we made use of rotational symmetry. 

Within the last part of \eq\nr{dx_exact}, a similar expansion
can be carried out with respect to $R$ in $1/(R-P)^2$, etc, but 
of course {\em not} in $1/(Q-R)^2$, since both variables
are of the same magnitude in this case. The challenge then is to deal 
with the angular integrals over the directions of $\vec{q}, \vec{r}$.
Letting $z=\vec{q}\cdot\vec{r}/(qr)$, 
$q_n = \sigma\, i q$, $r_n = \rho\, i r$, 
$\sigma=\pm$, $\rho=\pm$, 
the most difficult 
structures are the ones containing 
\be
 \frac{1}{(Q-R)^2} = \frac{1}{2qr(\sigma\rho - z)}
 \;.
\ee
The averaging denoted by $[...]_{Q,R}$ leads to 
\ba
 \fr14 \sum_{\sigma=\pm}\sum_{\rho=\pm}
 \frac{1}{\sigma\rho - z} & = &  \frac{z}{1-z^2} \;, \\ 
 \fr14 \sum_{\sigma=\pm}\sum_{\rho=\pm}
 \frac{\sigma}{\sigma\rho - z} & = & 
 \fr14 \sum_{\sigma=\pm}\sum_{\rho=\pm}
 \frac{\rho}{\sigma\rho - z} \; = \; 0 \;,  \\
 \fr14 \sum_{\sigma=\pm}\sum_{\rho=\pm}
 \frac{\sigma\rho}{\sigma\rho - z} & = &  \frac{1}{1-z^2} \;. 
\ea
Antisymmetry kills the structure linear in $z$, unless there 
is another angular variable appearing in the numerator. Fixing the 
directions of $\vec{p}$, $\vec{r}$ and integrating over those of 
$\vec{q}$, the latter case produces
\be
 \left\langle \frac{ \hat{\vec{q}}\cdot\hat{\vec{r}}\;\hat{\vec{q}} }
                   { 1 - (\hat{\vec{q}}\cdot\hat{\vec{r}})^2 } 
 \right\rangle_{\hat{\vec{q}}}
 = 
 \hat{\vec{r}}
 \left\langle 
    \frac{z^2}{1-z^2}          
 \right\rangle_{z}
 = 
  \frac{\hat{\vec{r}}}{D-4}\left\langle 1 \right\rangle 
 \;, 
\ee
where we made use of rotational symmetry
and the dimensionally 
regularized angular integration measure. 
A subsequent integral 
over the directions of $\vec{r}$ might contain
\be
 \left\langle ({\vec{p}}\cdot\hat{\vec{r}})^2 \right\rangle_{\hat{\vec{r}}}
 = \frac{\vec{p}^2}{D-1} 
 \left\langle 1 \right\rangle \;.
\ee
The case without any angular variable in the numerator yields
\ba
 \left\langle \frac{1}{1-z^2} \right\rangle = % & = &
  \frac{D-3}{D-4}\left\langle 1 \right\rangle \;.
\ea
Setting finally $D=4-2\epsilon$, we obtain the expansions
listed in appendix~A (\eqs\nr{Ja_res}--\nr{Ij_res}).

%%%%%%%%%%%%%%%%%%%%%%%%%%%% SECTION %%%%%%%%%%%%%%%%%%%%%%%%%%%%%%%%%%%
%
\section{Euclidean momentum-space correlators}
\la{se:mom}

Inserting into \eqs\nr{Gtheta_bare}, \nr{Gchi_bare}
the asymptotic expansions of the master sum-integrals 
from appendix~A as well as the bare
gauge coupling from \eq\nr{gB}; expanding in $\epsilon$;
and inserting 
$ 
 1 = \mu^{-2\epsilon} \bmu^{2\epsilon}
 \frac{\exp({\gammaE \epsilon})}{(4\pi)^\epsilon}
$, 
% where $\bmu$ is the scale parameter of the $\msbar$ scheme
% (the factors $\mu^{-2\epsilon}$ are not shown explicitly, having
% been redefined to be part of $g^2$ or the $\tilde G$'s),  
we can write the results in terms of the renormalized coupling
as follows:
\ba
 && \hspace*{-1cm} \frac{\tilde G_\theta(P)}
 {4 d_A c_\theta^2 g^4 \mu^{2\epsilon}}   \nn
 & = & 
 \frac{P^4}{(4\pi)^2}
 \biggl\{
 \biggl(\frac{\bmu}{P} \biggr)^{2\epsilon}
 \biggl[ \frac{1}{\epsilon} + 1 +... \biggr]
 \biggl( 1 - \frac{g^2 \Nc}{(4\pi)^2} \frac{22}{3\epsilon} \biggr)
 + 
 \frac{g^2 \Nc}{(4\pi)^2} 
 \biggl(\frac{\bmu}{P} \biggr)^{4\epsilon}
 \biggl[
  \frac{11}{3\epsilon^2} + \frac{95}{6\epsilon} +... 
 \biggr]
 \biggr\}
 \nn 
 & + & 
 \frac{8}{P^2} \biggl( \frac{p^2}{3} - p_n^2 \biggr)
 \biggl[
    1 + \frac{g^2\Nc}{(4\pi)^2}
  \biggl(
    \frac{22}{3} \ln\frac{\bmu^2}{P^2} + \frac{203}{18} 
  \biggr) 
 \biggr]
 \int_{\vec{q}} q\, \nB{}(q) 
 \nn 
 & - & 4 g^2 \Nc 
 \biggl[ \frac{3}{P^2} \biggl( \frac{p^2}{3} - p_n^2 \biggr) + 1 \biggr] 
  \int_{\vec{q},\vec{r}}
 \frac{\nB{}(q)}{q} \frac{\nB{}(r)}{r}
  + \rmO\Bigl(g^4,\frac{1}{P^2}\Bigr)
 \;, \la{theta_P} \\ 
%%%%%
%%%%%
  && \hspace*{-1cm} 
 \frac{\tilde G_\chi(P)}
      {-16 d_A c_\chi^2 g^4  \mu^{2\epsilon}}   \nn 
 & = & 
 \frac{P^4}{(4\pi)^2}
 \biggl\{
 \biggl(\frac{\bmu}{P} \biggr)^{2\epsilon}
 \biggl[ \frac{1}{\epsilon} - 1 +... \biggr]
 \biggl( 1 - \frac{g^2 \Nc}{(4\pi)^2} \frac{22}{3\epsilon} \biggr)
 + 
 \frac{g^2 \Nc}{(4\pi)^2} 
 \biggl(\frac{\bmu}{P} \biggr)^{4\epsilon}
 \biggl[
  \frac{11}{3\epsilon^2} + \frac{25}{2\epsilon} +... 
 \biggr]
 \biggr\}
 \nn 
 & + & 
 \frac{8}{P^2} \biggl( \frac{p^2}{3} - p_n^2 \biggr)
 \biggl[
    1 + \frac{g^2\Nc}{(4\pi)^2}
  \biggl(
    \frac{22}{3} \ln\frac{\bmu^2}{P^2} + \frac{347}{18} 
  \biggr) 
 \biggr]
 \int_{\vec{q}} q\, \nB{}(q) 
 \nn 
 & - & 4 g^2 \Nc 
 \biggl[ \frac{3}{P^2} \biggl( \frac{p^2}{3} - p_n^2 \biggr) - 1 \biggr] 
  \int_{\vec{q},\vec{r}}
 \frac{\nB{}(q)}{q} \frac{\nB{}(r)}{r}
  + \rmO\Bigl(g^4,\frac{1}{P^2}\Bigr)
 \;. \la{chi_P} 
\ea
Here and in the following, terms of $\rmO(\epsilon)$ have been 
omitted where irrelevant. 

The vacuum parts, i.e.\ the first rows of \eqs\nr{theta_P}, \nr{chi_P}, 
can be compared with ref.~\cite{old}. If we rewrite
$
 (\bmu/P)^{\alpha} = (\bmu e/P)^{\alpha}[1-\alpha +...]
$
and re-expand the square brackets, 
we can reproduce the results of ref.~\cite{old}. 

As far as the thermal parts go we note that, within the accuracy
of our computation, they can be expressed as 
\ba
% && \hspace*{-1cm}
 \frac{\Delta\tilde G_\theta(P)}{4 c_\theta^2 g^4}
% =  \nn
 & = & 
 \frac{3}{P^2} \biggl( \frac{p^2}{3} - p_n^2 \biggr)
 \biggl[
    1 + \frac{g^2\Nc}{(4\pi)^2}
  \biggl(
    \frac{22}{3} \ln\frac{\bmu^2}{P^2} + \frac{203}{18} 
  \biggr) 
 \biggr] \, (e+p)(T)
 \nn 
 & - & \frac{2}{ g^2 b_0}  
 \biggl[ 1 + g^2 b_0 
 \ln \frac{\bmu^2}{\zeta_\theta P^2}  \biggr]\, (e-3p)(T) 
  + \rmO\Bigl(g^4,\frac{1}{P^2}\Bigr)
 \;, \la{D_theta_P} \\ 
%%%%%
%%%%%
%  && \hspace*{-1cm} 
 \frac{\Delta \tilde G_\chi(P)}{-16 c_\chi^2 g^4} 
% = \nn 
 & = & 
 \frac{3}{P^2} \biggl( \frac{p^2}{3} - p_n^2 \biggr)
 \biggl[
    1 + \frac{g^2\Nc}{(4\pi)^2}
  \biggl(
    \frac{22}{3} \ln\frac{\bmu^2}{P^2} + \frac{347}{18} 
  \biggr) 
 \biggr]
 \, (e+p)(T)
 \nn 
 & + & \frac{2}{ g^2 b_0}  
 \biggl[ 1 + g^2 b_0 \ln \frac{\bmu^2}{\zeta_\chi P^2}  \biggr]\, (e-3p)(T)
  + \rmO\Bigl(g^4,\frac{1}{P^2}\Bigr)
 \;. \la{D_chi_P} 
\ea
Here we have identified the structures of \eqs\nr{eT}, \nr{em3pT}
in the result, realizing that the expressions
proportional to $(\frac{p^2}{3} - {\scriptstyle p_n^2})/P^2$ couple to the 
{\em traceless} part of the energy-momentum tensor, $\hat T_{\mu\nu}$, 
satisfying 
$
 \langle \hat T_{\mu\nu} \rangle = 
 \langle T_{00} \rangle 
 (\delta_{0 \mu }\delta_{0 \nu} -
  \delta_{i\mu}\delta_{i\nu}/3)
$, 
whereas the other terms couple to the trace part. 
In addition we have used renormalization group
invariance of $(e-3p)(T)$
as well as the theoretical expectation that the Wilson coefficients
should be independent of the ``soft scale'', $T$, 
to provisionally add to the results the logarithmic terms on the second rows; 
the coefficients $\zeta_\theta, \zeta_\chi$ 
next to $P^2$ remain undetermined, because 
fixing them would necessitate a 3-loop computation. 
The qualitative structure of \eq\nr{D_theta_P} agrees 
with that put forward in ref.~\cite{sch}.
(Note that the term proportional to $e+p$ does {\em not} appear
in classic vacuum studies like ref.~\cite{ope}, because it 
breaks Lorentz symmetry. In fact, expressed in another way, 
$(e+p)(T) = T s(T)$, where $s$ denotes the entropy density.)

%%%%%%%%%%%%%%%%%%%%%%%%%%% SECTION %%%%%%%%%%%%%%%%%%%%%%%%%%%%%%%%%%
%
\section{Short distances}
\la{se:r}

As a first concrete application, 
we consider equal-time correlators in configuration space, measured
recently in ref.~\cite{hbm_c}. This means that we need to inverse 
Fourier transform the correlators in \eq\nr{GP} in order to get back 
to the correlators of \eq\nr{Gx}. 
Since we are applying the inverse transform not 
to the full result but to an asymptotic expansion valid in 
the regime $P^2 \gg (\pi T)^2$, the inverse transform can (and must)
be taken at zero temperature (i.e.\ omitting terms like $\nB{}(p)$).
The master formula for an inverse Fourier transform in dimensional
regularization reads
\be
 \int_P \frac{e^{i P\cdot x}}{[P^2]^\alpha}
 = \frac{\Gamma(D/2 - \alpha)}{(\pi^\fr12 x)^D (x/2)^{-2\alpha}\Gamma(\alpha)}
 \;. \la{Fourier}
\ee
We recall from \eq\nr{def_P_x} that $r$ 
denotes a spatial separation in radial coordinates, 
$r = |\vec{x}|$, whereas 
the temporal separation is chosen to vanish 
as in ref.~\cite{hbm_c}, $\tau = 0$.

Employing \eq\nr{Fourier}, we obtain for the structures appearing 
in the vacuum parts (the first rows of \eqs\nr{theta_P}, \nr{chi_P})
the expansions
\ba
 \int_P e^{i P\cdot x}\, \frac{P^4}{(4\pi)^2} 
 \biggl( \frac{\bmu}{P} \biggr)^{2\epsilon} 
 & = &
 \frac{12 \mu^{-2\epsilon}}{\pi^4 r^8} (r\bmu)^{4\epsilon}
 \epsilon \biggl[1 + \epsilon
          \biggl( - \frac{31}{6} + 4 \gammaE - 4 \ln 2 \biggr)
  + \rmO(\epsilon^2)\biggr] 
 \;, \la{first} \\ 
 \int_P e^{i P\cdot x}\, \frac{P^4}{(4\pi)^2} 
 \biggl( \frac{\bmu}{P} \biggr)^{4\epsilon} 
 & = &
 \frac{24 \mu^{-2\epsilon}}{\pi^4 r^8} (r\bmu)^{6\epsilon}
 \epsilon \biggl[1 + \epsilon
          \biggl( - \frac{17}{2} + 6 \gammaE - 6 \ln 2 \biggr)
  + \rmO(\epsilon^2) \biggr] 
 \;.
\ea
Note that these are proportional to $\epsilon$, 
which is why we did not need to show terms of $\rmO(1)$ in 
\eqs\nr{theta_P}, \nr{chi_P}.
The structures multiplying $e+p$ in \eqs\nr{D_theta_P}, \nr{D_chi_P} yield
\ba
 \int_P e^{i P\cdot x} \frac{1}{P^2}\biggl(\frac{p^2}{3} - p_n^2 \biggr)
 & = & 
 -\frac{2}{3\pi^2 r^4}
 \;, \\ 
 \int_P e^{i P\cdot x} \frac{1}{P^2}\biggl(\frac{p^2}{3} - p_n^2 \biggr) 
 \ln \frac{\bmu^2}{P^2}
 & = & 
 -\frac{2}{3\pi^2 r^4}
  \biggl(2 \ln \frac{r \bmu e^{\gammaE}}{2} - \fr32 \biggr)
 \;. 
\ea
(To arrive at these it is helpful to consider the 
covariant integral 
$
 \int_P e^{i P\cdot x} P_\mu P_\nu / [P^2]^\alpha
$
first.)
Finally, the structures multiplying $e-3p$ in \eqs\nr{D_theta_P}, 
\nr{D_chi_P} contain kind of an ambiguity because, taken literally, 
the terms independent of $P$ yield $\delta^{(D)}(x)$, whereas 
according to \eq\nr{Fourier} we 
get nothing. % (after the limit $\alpha\to 0$). 
This ambiguity is of little significance, however, 
since the contact terms are of no interest at $r\neq 0$. 
In contrast, the terms with $\ln(\bmu^2/P^2)$ as an integrand 
yield a physical behaviour in dimensional regularization:
\be
 \int_P e^{i P\cdot x} 
 \ln \frac{\bmu^2}{P^2}
 = \lim_{\epsilon\to 0}
 \int_P e^{i P\cdot x} 
 \frac{1}{\epsilon}\biggl[
 \biggl( \frac{\bmu}{P} \biggr)^{2\epsilon} - 1
 \biggr]
 = \frac{1}{\pi^2 r^4}
 \;. \la{last}
\ee
Inserting \eqs\nr{first}--\nr{last} into the inverse Fourier 
transforms of \eqs\nr{theta_P}, \nr{chi_P}, we get
\ba
 \frac{G_\theta(r)}{4 c_\theta^2} 
 & = & 
 \frac{12d_A}{\pi^4 r^8}
 \,\gamma_\rmi{$\theta;\unit$}(r)
 \; - \;  
 \frac{2(e+p)}{\pi^2 r^4}
 \,\gamma_\rmi{$\theta;e+p$}(r)
 \; -  \;
 \frac{2(e-3p)}{\pi^2 r^4}
 \,\gamma_\rmi{$\theta;e-3p$}(r)
 \; + \; \rmO\biggl( \frac{T^6}{r^2} \biggr) 
 \;, \nn \la{G_theta_r} \\ 
%%%%%%%%%%%%%%%%%%%%%%%%%%%%%%%%%%%%%%%%%%%%%%%%%%%
%%%%%%%%%%%%%%%%%%%%%%%%%%%%%%%%%%%%%%%%%%%%%%%%%%%
 \frac{G_\chi(r)}{-16 c_\chi^2 } 
 & = & 
 \frac{12d_A}{\pi^4 r^8}
 \,\gamma_\rmi{$\chi;\unit$}(r)
 \; - \;
 \frac{2(e+p)}{\pi^2 r^4}
 \,\gamma_\rmi{$\chi;e+p$}(r)
 \; + \; 
 \frac{2(e-3p)}{\pi^2 r^4}
 \,\gamma_\rmi{$\chi;e-3p$}(r)
 \; + \; \rmO\biggl( \frac{T^6}{r^2} \biggr) 
 \;, \nn \la{G_chi_r}
\ea
where
\ba
 \gamma_\rmi{$\theta;\unit$}(r) & = & 
  g^4 + \frac{g^6\Nc}{(4\pi)^2}
 \biggl( \frac{44}{3}\ln \frac{r \bmu e^{\gammaE}}{2} - \fr1{9} \biggr)
 + \rmO(g^8) 
 \;, \la{gamma_theta_unit} \\ 
 \gamma_\rmi{$\theta;e+p$}(r) & = & 
  g^4 + \frac{g^6\Nc}{(4\pi)^2}
 \biggl( \frac{44}{3}\ln \frac{r \bmu e^{\gammaE}}{2} + \fr5{18} \biggr) 
 + \rmO(g^8) 
 \;, \la{gamma_theta_ep} \\[2mm] 
 \gamma_\rmi{$\theta;e-3p$}(r) & = & 
  g^4 + \rmO(g^6) 
 \;, \la{gamma_theta_emtp} \\[2mm]
 \gamma_\rmi{$\chi;\unit$}(r) & = & 
  g^4 + \frac{g^6\Nc}{(4\pi)^2}
 \biggl( \frac{44}{3}\ln \frac{r \bmu e^{\gammaE}}{2} + \fr{71}{9} \biggr) 
 + \rmO(g^8) 
 \;, \la{gamma_chi_unit} \\ 
 \,\gamma_\rmi{$\chi;e+p$}(r) 
 & = &  
 g^4 + \frac{g^6\Nc}{(4\pi)^2}
 \biggl( \frac{44}{3}\ln \frac{r \bmu e^{\gammaE}}{2} + \fr{149}{18} \biggr) 
 + \rmO(g^8) 
 \;, \la{gamma_chi_ep} \\[2mm] 
 \gamma_\rmi{$\chi;e-3p$}(r) & = & 
  g^4 + \rmO(g^6) 
 \;. \la{gamma_chi_emtp}
\ea

%%%%%%%%%%%%%%%%%%%%%%%%%%%%%%%%% FIGURE %%%%%%%%%%%%%%%%%%%%%%%%%%%%%%%%%
\begin{figure}[t]

%\vspace*{-3cm}

\centerline{%
 \epsfysize=9.0cm\epsfbox{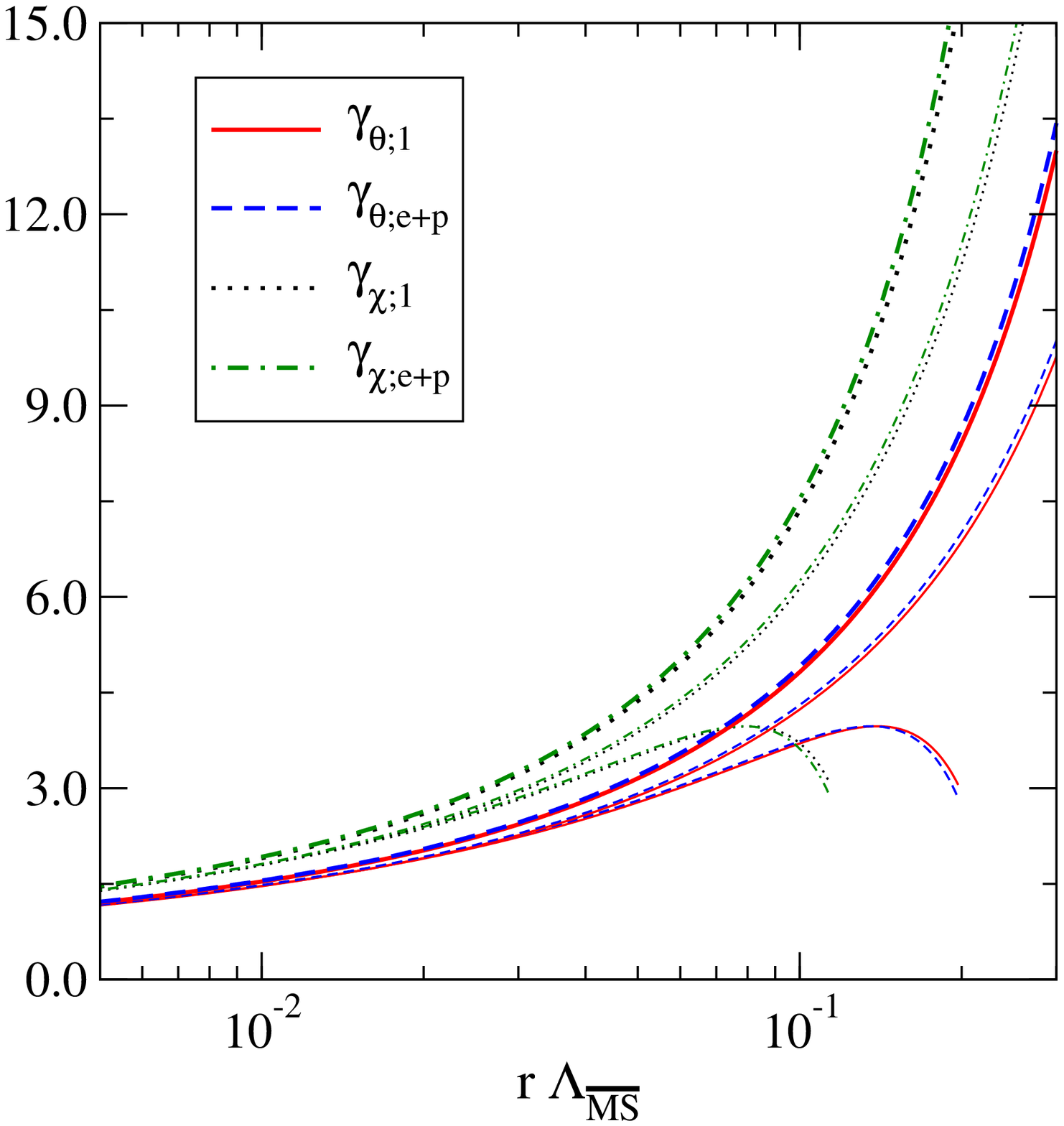}%
}

\caption[a]{\small 
  Numerical estimates of the Wilson coefficients 
  $\gamma_\rmi{$\theta;\unit$}$,
  $\gamma_\rmi{$\theta;e+p$}$,
  $\gamma_\rmi{$\chi;\unit$}$, 
  $\gamma_\rmi{$\chi;e+p$}$, 
  cf.\ \eqs\nr{gamma_theta_unit}, \nr{gamma_theta_ep}, 
  \nr{gamma_chi_unit}, \nr{gamma_chi_ep}, respectively.
  Thick lines correspond to choosing $\bmu$ such that the next-to-leading
  order correction vanishes (we refer to this value as $\bmu_\rmi{opt}$); 
  thin lines correspond to 
  $\bmu = 0.5 \bmu_\rmi{opt}$ or
  $\bmu = 2.0 \bmu_\rmi{opt}$
  (the dependence on $\bmu$ is non-monotonic). 
  The gauge coupling is solved from the 2-loop renormalization 
  group equation, and
  $
   \Lambdamsbar \equiv \lim_{\bmu\to\infty}
   \bmu \bigl[ b_0 g^2 \bigr] ^{-b_1/2 b_0^2}
   \exp \bigl[ -\frac{1}{2 b_0 g^2}\bigr]
  $. 
  Particularly in the $\chi$-channel
  reasonable apparent convergence is observed only 
  at extremely short distances. 
 }
\la{fig:rdep}
\end{figure}
%%%%%%%%%%%%%%%%%%%%%%%%%%%%%%%%%%%%%%%%%%%%%%%%%%%%%%%%%%%%%%%%%%%%%%%%%%%

The results 
of \eqs\nr{G_theta_r}, \nr{G_chi_r}
can be compared with those given in refs.~\cite{hbm_b,hbm_c}.
Inserting $c_\theta$ from \eq\nr{c_theta},  our leading order 
results for the vacuum and trace anomaly parts of $G_\theta$ 
agree with ref.~\cite{hbm_b}. In contrast, we find a coefficient of 
the term proportional to $e+p$ to be larger by a factor 2 
than in ref.~\cite{hbm_b}.\footnote{%
 However the correct coefficient (in momentum space)
 can be found in the more recent ref.~\cite{hbm_d}.
 } 
We remark that this coefficient appears both at $\rmO(g^4)$ and 
$\rmO(g^6)$, and we get contributions at both orders which combine 
to produce the correct renormalization group invariant structure. 
In addition, the leading order result can 
even be worked out exactly: omitting terms that 
vanish in dimensional regularization, we get
\be
 \frac{G_\theta(r)}{4 c_\theta^2 } 
 = 
 \frac{G_\chi(r)}{-16 c_\chi^2 } 
 = 
 4 d_A g^4
 \{ \partial_\mu \partial_\nu \Delta(x)
    \partial_\mu \partial_\nu \Delta(x) \}_{\tau = 0}  
 + \rmO(g^6)
 \;, 
\ee
where $\Delta(x)$ is the scalar propagator in coordinate space, 
\be
 \Delta(x) 
 \equiv
 \Tint{P} \frac{e^{i P\cdot x}}{P^2}
 = \frac{T}{4\pi r} \re \Bigl[\mathop{\mbox{coth}}(\pi T(r + i \tau)) \Bigr]
 + \rmO(\epsilon) \;.
\ee
Carrying out the derivatives in radial coordinates 
[$\partial_\mu\partial_\nu \Delta \partial_\mu\partial_\nu \Delta 
 = \partial_\tau^2 \Delta \partial_\tau^2 \Delta  
 + 2 \partial_\tau \partial_r \Delta \partial_\tau \partial_r \Delta 
 + (2/r^2) \partial_r \Delta \partial_r \Delta
 + \partial_r^2 \Delta \partial_r^2 \Delta
$] and setting $\tau = 0$ in the end, we obtain an 
elementary if complicated expression, whose expansion in a small $rT$
yields
\be 
  \frac{G_\theta(r)}{4 c_\theta^2 } 
 = 
 \frac{G_\chi(r)}{-16 c_\chi^2 } 
 = 
 4 d_A g^4 \biggl\{ 
   \frac{3}{\pi^4 r^8} - \frac{2 T^4}{45 r^4} + \frac{8\pi^2 T^6}{315 r^2}
 + \rmO(1)
 \biggr\} + \rmO(g^6)
 \;. 
\ee
Identifying $e+p$ from \eqs\nr{eT}, \nr{intrep} 
we reproduce the two leading terms 
of \eq\nr{G_theta_r}. (We can also read from here that 
the dimensionless expansion parameter is $\sim (r \pi T)^2$.)

Concerning $G_\chi$, it is stated in
ref.~\cite{hbm_c} that it has the same Operator Production Expansion
as $G_\theta$, whereas we find a different sign for the term 
proportional to the trace anomaly, $e-3p$. Our finding 
is consistent with ref.~\cite{ope} (cf.\ \eq(3.14) there):  
if the vacuum terms are normalized to be equal in the two cases, 
as has been done in \eqs\nr{G_theta_r}, \nr{G_chi_r}, then the trace 
anomaly terms come with opposite signs, and cancel in the sum.
We note, furthermore, that the numerical
results of ref.~\cite{hbm_c} show a very different 
short-distance behaviour of the thermal parts of  
the two correlators (cf.\ \fig5), which could also be a reflection  
of the fact that  $- G_\chi$ gets a positive 
contribution from the trace anomaly at short distances whereas 
$G_\theta$ gets a negative contribution. (It must be stressed, though, 
that if the Wilson coefficients of the structures proportional to $e+p$
and $e-3p$ were the same, as is the case at leading order, then 
the positive contribution from $e-3p$ could not
overcome the negative contribution from $e+p$ in $-G_\chi$, 
given that $-(e+p)+(e-3p) = -4p < 0$.)

We conclude by noting that the Wilson coefficients in  
\eqs\nr{gamma_theta_unit}, \nr{gamma_theta_ep}, 
\nr{gamma_chi_unit}, \nr{gamma_chi_ep} go beyond
the accuracy of the analysis in refs.~\cite{hbm_b,hbm_c}. 
With next-to-leading order corrections available, we should be in a position
to estimate the coefficients numerically; this has been 
attempted in \fig\ref{fig:rdep}. Any sort of  apparent convergence 
is only observed at very short distances, particularly
in the $\chi$-channel; this unfortunate 
fact may not be totally unexpected~\cite{david}. In any case, 
it can be seen that the various Wilson coefficients 
could be numerically quite different even though 
they agree at leading order. 
In principle it would be nice
to also determine the Wilson coefficients related to the trace 
anomaly terms (\eqs\nr{gamma_theta_emtp}, \nr{gamma_chi_emtp})
at next-to-leading order but 
this would require a 3- or 4-loop computation. 
(Since the operator yielding $e-3p$ is Lorentz invariant 
and the Wilson coefficients should be $T$-independent, 
it might be possible to extract these
coefficients from purely vacuum computations, however we 
have not unearthed literature where this would have been achieved.)

%%%%%%%%%%%%%%%%%%%%%%%%%%%%% SECTION %%%%%%%%%%%%%%%%%%%%%%%%%%%%%%%%%%%
%
\section{Large frequencies}
\la{se:w}

We next turn to an ``opposite'' limit, that of large frequencies
but vanishing spatial momenta: $P = (p_n,\vec{0})$. The results
of \eqs\nr{theta_P}--\nr{D_chi_P} remain valid as a starting point. 
Furthermore, 
it has been argued in ref.~\cite{sch} that the asymptotic expansions
can be analytically continued to Minkowski signature, 
$p_n \to -i [\omega + i 0^+]$, 
extracting thereby the spectral functions, 
$\rho(\omega) = \im \tilde G(-i [\omega + i 0^+])$, 
even though this requires crossing the 
light-cone, $P^2=0$, on which the asymptotic expansions are certainly 
{\em not} valid. In any case, within perturbation theory, 
the procedure should surely 
be justified, since we could imagine carrying out the analytic continuation
before the asymptotic expansion in each individual master sum-integral, 
and expanding only subsequently in a large $\omega/\pi T$.

Let us stress that we assume the analytic continuation to be carried
out {\em in the presence of an ultraviolet regulator} 
for spatial momenta.
The regulator is removed (i.e.\ the limit $\epsilon\to 0$ is taken) only
{\em after the analytic continuation}. 

With these qualifications 
the structures in \eqs\nr{theta_P}--\nr{D_chi_P} yield
(at $\vec{p} = \vec{0}$ and $\omega > 0$)
\ba
 \frac{P^4}{(4\pi)^2}
 \biggl(\frac{\bmu}{P} \biggr)^{2\epsilon}
 & \longrightarrow &
 \frac{\omega^4}{(4\pi)^2}\, \epsilon\pi 
 \biggl[1 + \epsilon\ln\frac{\bmu^2}{\omega^2} \biggr] 
 \;, \\
 \frac{P^4}{(4\pi)^2}
 \biggl(\frac{\bmu}{P} \biggr)^{4\epsilon}
 & \longrightarrow &
 \frac{\omega^4}{(4\pi)^2}\, 2 \epsilon\pi 
 \biggl[1 + 2 \epsilon\ln\frac{\bmu^2}{\omega^2} \biggr] 
 \;, \\ % \ea \ba  
 \frac{1}{P^2} \biggl( \frac{p^2}{3} - p_n^2 \biggr)
 \ln\frac{\bmu^2}{P^2} 
 & \longrightarrow &
 -\pi
 \;, \\ 
 \ln\frac{\bmu^2}{P^2} 
 & \longrightarrow & 
 \pi 
 \;. 
\ea
Thereby we get the spectral functions 
\ba
 \frac{\rho_\theta(\omega)}{4c_\theta^2 \pi}
 & = & 
 \frac{d_A \omega^4}{(4\pi)^2}
 \,\tilde\gamma_\rmi{$\theta;\unit$}(\omega)
 \; - \; {2(e+p)} 
 \,\tilde\gamma_\rmi{$\theta;e+p$}(\omega)
 \; - \; 2 (e-3p)
 \,\tilde\gamma_\rmi{$\theta;e-3p$}(\omega)
 \; + \; \rmO\biggl( \frac{T^6}{\omega^2} \biggr)
 \;, \nn \la{rho_theta} \\ 
%%%%%%%%%%%%%%%%%%%%%%%
 \frac{\rho_\chi(\omega)}{-16c_\chi^2  \pi}
 & = & 
 \frac{d_A \omega^4}{(4\pi)^2}
 \,\tilde\gamma_\rmi{$\chi;\unit$}(\omega)
 \; - \; {2(e+p)} 
 \,\tilde\gamma_\rmi{$\chi;e+p$}(\omega)
 \; + \; 2 (e-3p)
 \,\tilde\gamma_\rmi{$\chi;e-3p$}(\omega)
 \; + \; \rmO\biggl( \frac{T^6}{\omega^2} \biggr)
 \;, \nn \la{rho_chi}
\ea
where
\ba
 \tilde\gamma_\rmi{$\theta;\unit$}(\omega) & = & 
 g^4 + \frac{g^6\Nc}{(4\pi)^2}
 \biggl( \frac{22}{3}\ln\frac{\bmu^2}{\omega^2} + \frac{73}{3} \biggr) 
 + \rmO(g^8)
 \;, \la{tgamma_theta_unit} \\ 
 \tilde\gamma_\rmi{$\theta;e+p$}(\omega) & = & 
 \frac{11g^6 \Nc}{(4\pi)^2} + \rmO(g^8)
 \;, \la{tgamma_theta_ep} \\[2mm] 
 \tilde\gamma_\rmi{$\theta;e-3p$}(\omega) & = & 
 g^4 + \rmO(g^6)
 \;, \la{tgamma_theta_emtp} \\[2mm] 
 \tilde\gamma_\rmi{$\chi;\unit$}(\omega) & = & 
 g^4 + \frac{g^6\Nc}{(4\pi)^2}
 \biggl( \frac{22}{3}\ln\frac{\bmu^2}{\omega^2} + \frac{97}{3} \biggr) 
 + \rmO(g^8) 
 \;, \la{tgamma_chi_unit} \\ 
 \tilde\gamma_\rmi{$\chi;e+p$}(\omega) & = & 
 \frac{11g^6 \Nc}{(4\pi)^2} + \rmO(g^8)
 \;, \la{tgamma_chi_ep} \\[2mm] 
 \tilde\gamma_\rmi{$\chi;e-3p$}(\omega) & = & 
 g^4 + \rmO(g^6)
 \;. \la{tgamma_chi_emtp}
\ea

After inserting $c_\theta$ from \eq\nr{c_theta} the leading-order 
term of $\tilde\gamma_\rmi{$\theta;\unit$}$ 
agrees with ref.~\cite{hbm_a}, but we can 
now add to that result the first correction. Similarly, the leading-order
term of $\tilde\gamma_\rmi{$\chi;\unit$}$ 
agrees with a result given in ref.~\cite{hbm_c}, 
but we can add a correction.
The coefficients 
$
\tilde\gamma_\rmi{$\theta;e+p$}, 
\tilde\gamma_\rmi{$\theta;e-3p$}
$
agree with ref.~\cite{sch}, 
if the coefficient $C$ introduced there is set to $C=1$. The results for 
$
\tilde\gamma_\rmi{$\theta;e+p$}, 
\tilde\gamma_\rmi{$\theta;e-3p$}
$
have more recently been reproduced in ref.~\cite{hbm_d}. We are not aware
of analogous results in the literature for 
$
\tilde\gamma_\rmi{$\chi;e+p$}, 
\tilde\gamma_\rmi{$\chi;e-3p$}
$.

%%%%%%%%%%%%%%%%%%%%%%%%%%%%%%%%% FIGURE %%%%%%%%%%%%%%%%%%%%%%%%%%%%%%%%%
\begin{figure}[t]

%\vspace*{-3cm}

\centerline{%
 \epsfysize=9.0cm\epsfbox{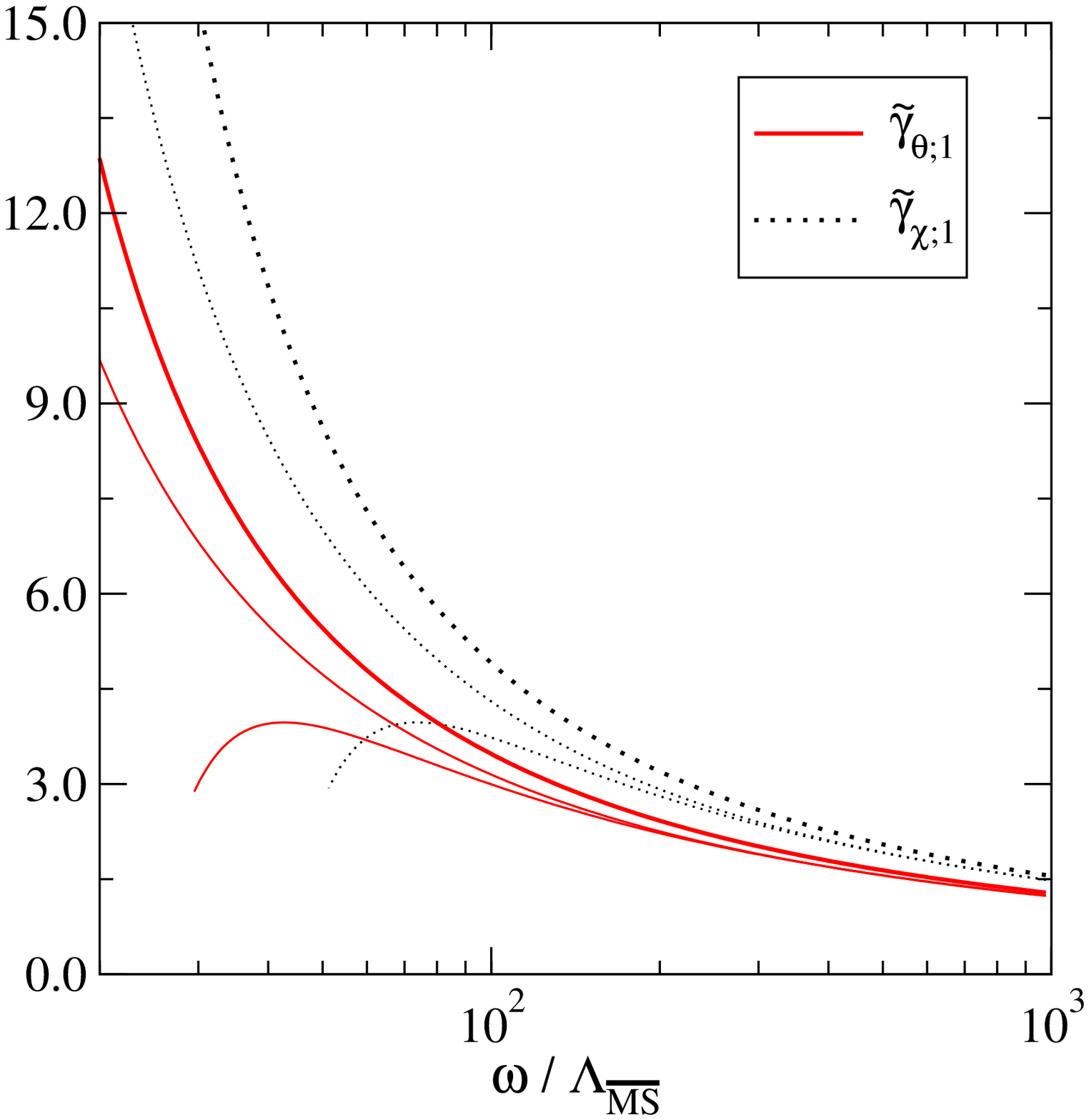}%
}

\caption[a]{\small 
  Numerical estimates of the Wilson coefficients 
  $\tilde\gamma_\rmi{$\theta;\unit$}$,
  $\tilde\gamma_\rmi{$\chi;\unit$}$, 
  cf.\ \eqs\nr{tgamma_theta_unit}, 
  \nr{tgamma_chi_unit}, respectively.
  Unspecified conventions are as in \fig\ref{fig:rdep}. 
  Reasonable apparent convergence is observed only 
  at extremely large frequencies, 
  $\omega\gsim 50\, \Lambdamsbar$ ($\tilde\gamma_\rmi{$\theta;\unit$}$)
  or  $\omega\gsim 80\, \Lambdamsbar$ ($\tilde\gamma_\rmi{$\chi;\unit$}$). 
 }
\la{fig:wdep}
\end{figure}
%%%%%%%%%%%%%%%%%%%%%%%%%%%%%%%%%%%%%%%%%%%%%%%%%%%%%%%%%%%%%%%%%%%%%%%%%%%

The coefficients 
$\tilde\gamma_\rmi{$\theta;\unit$}$,
$\tilde\gamma_\rmi{$\chi;\unit$}$, 
for which next-to-leading order values are given
in \eqs\nr{tgamma_theta_unit}, \nr{tgamma_chi_unit}, 
are estimated numerically in \fig\ref{fig:wdep}.
Like in \se\ref{se:r}, reasonable apparent convergence is 
observed only extremely deep into the ultraviolet regime. 

We note that 
$
 \tilde\gamma_\rmi{$\theta;e-3p$}, 
 \tilde\gamma_\rmi{$\chi;e-3p$}
$
(\eqs\nr{tgamma_theta_emtp}, \nr{tgamma_chi_emtp})
contain second and higher powers of $g^2$, so that they should 
gradually decrease with $\omega$ as dictated by asymptotic freedom
(in analogy with \fig\ref{fig:wdep}). On the other hand, 
if similar results were available in the ``shear channel'', 
i.e.\ for 2-point correlators of the traceless part 
of the energy-momentum tensor, 
then the terms proportional to $e-3p$ would allow to fix an unknown
constant, denoted by $D$ in ref.~\cite{sch}, which is argued to be 
``saturated'' at that level~\cite{sch}. In that case there is no 
$g^4$ multiplying it, so the result is a constant. This implies
that such terms require a special treatment in the context of sum rules,
cf.\ \se\ref{se:sum}. 
The importance of determining such terms has recently been 
pointed out also in the context of a lattice investigation~\cite{hbm_e}.
Unfortunately, determining the 2-point correlators in the shear channel
is technically
more demanding than the present analysis because of the more cumbersome
Lorentz structures appearing in the numerator.

%%%%%%%%%%%%%%%%%%%%%%%%%%%%% SECTION %%%%%%%%%%%%%%%%%%%%%%%%%%%%%%%%%%%%
%
\section{On sum rules}
\la{se:sum}

Starting from the spectral representation of a Euclidean correlator, 
\be
 \tilde G (p_n) = \int_{-\infty}^{\infty} \frac{{\rm d}\omega}{\pi}
 \frac{\rho(\omega)}{\omega - i p_n}
 \;, \la{spec_rep}
\ee
which is generally valid in the presence of an ultraviolet regulator
for spatial momenta,\footnote{%
 Equation \nr{spec_rep} can be derived for instance by Fourier 
 transforming the known non-perturbative relation 
 in~\eq\nr{int_rel},
 which is unproblematic if $G(\tau)$ does not diverge 
 at small distances. 
 } 
and setting $p_n\to 0$, we obtain 
\be
 \int_0^\beta \! {\rm d}\tau \, G(\tau) =
 \int_{-\infty}^{\infty} \frac{{\rm d}\omega}{\pi}
 \frac{\rho(\omega)}{\omega}
 \;.  \la{srule}
\ee
Adding the dependence on spatial coordinates (which were suppressed above), 
we thus obtain a ``sum rule'', relating an integral over the spectral function
to a Euclidean ``susceptibility''. 

Now, in practice, various complications can arise in the application
of sum rules. One of them is of an {\em infrared} type, and particularly
relevant for the ``bulk channel'' (our correlator $G_\theta$): indeed
there must be a term $\propto\omega\,\delta(\omega)$ in $\rho_\theta$ 
because, if we set $\vec{p}\to\vec{0}$ from the outset, the operator 
couples to a conserved charge $\int_{\vec{x}} T_{00}$. 
Such a term does yield a contribution to both sides of the sum rule, 
however not to the {\em transport coefficient} of interest 
($\lim_{\omega\to 0^+} \rho_\theta(\omega)/\omega$), 
so it would be wise to subtract it. In ref.~\cite{hbm_d}, 
this was achieved by defining a modified spectral function $\Delta \rho_*$ 
which has no $\omega\,\delta(\omega)$ but nevertheless yields the same 
transport coefficient. If the problematic term is subtracted from
the right-hand side, it must also be subtracted from 
the left-hand side, and this yields a modified sum rule 
(cf.\ appendix~C of ref.~\cite{rs}). However 
this extra contribution is of $\rmO(g^8)$ and thus 
not directly visible in our $\rmO(g^6)$ result. 

Another possible complication, of an {\em ultraviolet} type,  
arises if we wish to remove the ultraviolet
regularization {\em before} applying the sum rule. For instance,
the vacuum part then grows as $\rho \propto \omega^4\sign(\omega)$, 
and both sides
of \eq\nr{srule} are ill-defined. The problem becomes less severe if the 
vacuum parts are subtracted from both sides of \eq\nr{srule}, imposing
a relation between $\Delta G$ and $\Delta \rho$ instead. However, 
even such a subtraction may not be sufficient to remove a constant part, 
as is argued to be the case in the shear channel~\cite{rs,sch}; 
then \eq\nr{spec_rep} needs to contain a supplementary 
``contact term'' on the right-hand side.

We now return to the case at hand, where only the former problem
should be relevant~\cite{sch,hbm_d}.
The left-hand side of the sum rule  
reads $\tilde G_\theta(0)$, $\tilde G_\chi(0)$ in our notation. 
Inspecting the master sum-integrals in \eqs\nr{Ja}--\nr{Ij}, 
most are seen to vanish in the limit $P\to 0$; in fact only 
$\mathcal{I}_\rmi{a}$ and $\mathcal{I}_\rmi{i}$
give a contribution, and the latter can 
be reduced to the former through \eq\nr{relation}.
Inserting these values into \eqs\nr{Gtheta_bare}, \nr{Gchi_bare}
we obtain %, up to $\rmO(g^6)$, 
\ba
 \tilde G_\theta(0) & = &  
 - 8 d_A c_\theta^2 g^6 \Nc (D-2)^2 \, \mathcal{I}_\rmi{a}
 = - 8 d_A b_0^2 g^6 \Nc 
   \int_{\vec{q},\vec{r}}
   \frac{\nB{}(q)}{q} \frac{\nB{}(r)}{r}
 \;, \la{sum_theta}
 \\ 
 \tilde G_\chi(0) & = &  0
 \;,  \la{sum_chi}
\ea
where in the final step we made use of \eqs\nr{c_theta}, \nr{Ia_res}.
Equation~\nr{sum_theta} agrees with \eq\nr{Tem3pT} and therefore 
confirms the sum rule of ref.~\cite{ellis}, obtained also directly 
in lattice regularization~\cite{hbm_f}, up to $\rmO(g^6)$. 
Unfortunately, as mentioned, the infrared ambiguities discussed 
in refs.~\cite{rs,hbm_d} are of $\rmO(g^8)$ and thus 
beyond the resolution of our computation. 

%%%%%%%%%%%%%%%%%%%%%%%%%%%% SECTION %%%%%%%%%%%%%%%%%%%%%%%%%%%%%%%%%%%%%
%
\section{Summary and outlook}
\la{se:concl}

Following ref.~\cite{hbm_c} and other recent works, 
we have analyzed the ultraviolet asymptotics
of 2-point correlation functions of the trace of the energy-momentum tensor 
and of the topological charge density in finite-temperature Yang-Mills theory. 
The tool has been dimensionally regularized perturbation 
theory, pursued up to 2-loop order. We have considered both  
Euclidean (operators separated by a purely spatial separation; \se\ref{se:r}) 
and Minkowskian correlators 
(operators separated by a Euclidean timelike separation,
subsequently Fourier transformed and 
analytically continued; \se\ref{se:w}).

Through our analysis, we have confirmed a number of expressions 
in recent literature, but simultaneously also refined the determination of 
the corresponding Wilson coefficients, particularly in the Euclidean domain
(\eqs\nr{gamma_theta_unit}, \nr{gamma_theta_ep}, \nr{gamma_chi_unit}, 
\nr{gamma_chi_ep}). 
In this case, we have also identified some inaccuracies in the literature;
on the qualitative level, our expressions appear to be in a somewhat better 
accordance with lattice data than the earlier ones, because they show
a qualitative difference between the two channels, with the terms 
proportional to the thermal part of the 
trace anomaly, $(e-3p)(T)$, coming with opposite signs. 
Unfortunately the perturbative series for the Wilson coefficients show 
slow convergence and it appears uncertain whether, even with the correct 
values inserted, they can lead to a  quantitative 
agreement with lattice data. 
(The problem is worse in the case of the topological 
charge density correlator, 
cf.\ \fig\ref{fig:rdep}.)

As an outlook, we believe that our results 
can be refined in a number of ways, some of them 
with reasonable prospects for phenomenological success. 
First of all, 
it would be interesting to determine the full $r$-dependence
of the correlators up to distances $r\sim 1/(\pi T)$ rather 
than the asymptotic expansions at $r\ll 1/(\pi T)$ as in the present study. 
This would increase the range of applicability of the results and  
might simultaneously improve on the convergence of the weak-coupling 
expansion, both aspects facilitating a comparison
with lattice data {\em \`a la} ref.~\cite{hbm_c}.  However, we would 
suggest carrying out the computation 
after averaging over the time separation, $\tau$, 
rather than setting it to zero, because 
this makes the correlators more analogous to the ones 
encountered in the context of heavy quarkonium 
physics~\cite{singlet},
a problem of actual phenomenological significance.
In any case, \eqs\nr{Gtheta_bare}, \nr{Gchi_bare} can serve
as starting points for such investigations; furthermore, 
the asymptotic results for the time-averaged correlators 
can still be obtained from \eqs\nr{theta_P}--\nr{D_chi_P}, simply 
by replacing the inverse Fourier transforms in \eq\nr{Fourier} by
three-dimensional ones.   

Second, it would be interesting to determine the full $\omega$-dependence 
of the spectral functions down to frequencies $\omega\sim \pi T$, rather 
than carrying out an asymptotic expansion at $\omega\gg \pi T$ as 
in the present study because, as mentioned around \eq\nr{int_rel}, such 
information may be helpful in connection with lattice extractions of 
the corresponding transport coefficients. 
The full frequency dependence is believed to be quite non-trivial 
and display e.g.\ a large cancellation of terms of $\rmO(g^4T^4)$~\cite{ms,sch}.
Once more, \eqs\nr{Gtheta_bare}, \nr{Gchi_bare} can serve
as starting points for this exercise; moreover the asymptotic expansions
in \eqs\nr{rho_theta}, \nr{rho_chi} could provide 
for useful crosschecks on the final results.

Third, for theoretical completeness, 
it would be interesting to determine the Wilson 
coefficients more accurately, particularly the missing $\rmO(g^6)$
terms in \eqs\nr{gamma_theta_emtp}, \nr{gamma_chi_emtp}, because this would
allow us to estimate the renormalization scale relevant for 
the gauge coupling multiplying the trace anomaly. (Nevertheless, 
because of the reservations mentioned in the present study in 
the case of the other Wilson coefficients, it is not clear to us whether 
this could lead to a practically successful comparison 
with the lattice data of ref.~\cite{hbm_c}.)
Similarly, of course, it would in principle be nice to know all 
next-to-leading order corrections in the Minkowskian domain, 
relevant for \eqs\nr{tgamma_theta_ep}, \nr{tgamma_theta_emtp}, 
\nr{tgamma_chi_ep}, \nr{tgamma_chi_emtp}.
All of these challenges necessitate at least 3-loop computations. 

Fourth, it would be interesting to determine the dependence of the 
various Wilson coefficients on the number 
of massless quark flavours, $\Nf$, not least because this offers 
a further crosscheck on the applicability of the Operator Production 
Expansion framework in the finite-temperature context. 
Partial results (including only ``sea'' quarks), and the problems 
encountered in this computation, are discussed in appendix~C. 

Finally, it would be interesting to 
repeat the study in the ``shear'' channel, i.e.\ for 
the traceless part of the energy-momentum tensor. 
The general techniques of \se\ref{se:method}, as well as 
some of the actual sum-integrals and asymptotic expansions
encountered in the present study, might play a role in that 
context as well. 

We hope to return to some of these topics in future work.

%%%%%%%%%%%%%%%%%%%%%%%%% SECTION %%%%%%%%%%%%%%%%%%%%%%%%%%%%%%%%%%%%%
%
\section*{Acknowledgements}

We thank H.B.~Meyer for useful correspondence.
M.V.\ was supported by the Academy of Finland, 
contract no.\ 128792, and 
A.V.\ by the Sofja Kovalevskaja program 
of the Alexander von Humboldt foundation.

\newpage

%%%%%%%%%%%%%%%%%%%%%%% APPENDIX %%%%%%%%%%%%%%%%%%%%%%%%%%%%%%%%%%%
%
\appendix
\renewcommand{\thesection}{Appendix~\Alph{section}}
\renewcommand{\thesubsection}{\Alph{section}.\arabic{subsection}}
\renewcommand{\theequation}{\Alph{section}.\arabic{equation}}

%%%%%%%%%%%%%%%%%%%%%%%%%%%%%% SECTION %%%%%%%%%%%%%%%%%%%%%%%%%%%%%%%%%
%
\section{Bosonic master sum-integrals}

The bosonic ``master'' sum-integrals appearing in our computation are defined as 
\ba
 \mathcal{J}_\rmi{a} & \equiv & 
 \Tint{Q} \frac{P^2}{Q^2}
 \;, \la{Ja} \\ 
 \mathcal{J}_\rmi{b} & \equiv & 
 \Tint{Q} \frac{P^4}{Q^2(Q-P)^2}
 \;, \la{Jb} \\ 
 \mathcal{I}_\rmi{a} & \equiv & 
 \Tint{Q,R} \frac{1}{Q^2R^2}
 \;, \\ 
 \mathcal{I}_\rmi{b} & \equiv & 
 \Tint{Q,R} \frac{P^2}{Q^2R^2(R-P)^2}
 \;, \\ 
 \mathcal{I}_\rmi{c} & \equiv & 
 \Tint{Q,R} \frac{P^2}{Q^2R^4}
 \;, \\ 
 \mathcal{I}_\rmi{d} & \equiv & 
 \Tint{Q,R} \frac{P^4}{Q^2R^4(R-P)^2}
 \;, \\ 
 \mathcal{I}_\rmi{e} & \equiv & 
 \Tint{Q,R} \frac{P^2}{Q^2R^2(Q-R)^2}
 \;, \\ 
 \mathcal{I}_\rmi{f} & \equiv & 
 \Tint{Q,R} \frac{P^2}{Q^2(Q-R)^2(R-P)^2}
 \;, \\ 
 \mathcal{I}_\rmi{g} & \equiv & 
 \Tint{Q,R} \frac{P^4}{Q^2(Q-P)^2R^2(R-P)^2}
 \;, \\ 
 \mathcal{I}_\rmi{h} & \equiv & 
 \Tint{Q,R} \frac{P^4}{Q^2R^2(Q-R)^2(R-P)^2}
 \;, \la{def_Ih} \\ 
 \mathcal{I}_\rmi{i} & \equiv & 
 \Tint{Q,R} \frac{(Q-P)^4}{Q^2R^2(Q-R)^2(R-P)^2}
 \;, \\ 
 \mathcal{I}_\rmi{i'} & \equiv & 
 \Tint{Q,R} \frac{4(Q\cdot P)^2}{Q^2R^2(Q-R)^2(R-P)^2}
 \;, \\ 
 \mathcal{I}_\rmi{j} & \equiv & 
 \Tint{Q,R} \frac{P^6}{Q^2R^2(Q-R)^2(Q-P)^2(R-P)^2}
 \;. \la{Ij}
\ea
In fact there is some redundancy here, because the following relation
can easily be established through changes of integration variables: 
\be
 \mathcal{I}_\rmi{i}  =  
% \Tint{Q,R} \frac{(Q-P)^4}{Q^2R^2(Q-R)^2(R-P)^2}
 \mathcal{I}_\rmi{a} + \mathcal{I}_\rmi{e} - \mathcal{I}_\rmi{f}
 + \mathcal{I}_\rmi{i'}
 \;. \la{relation}
\ee

Expanding in a small $(\pi T)^2/P^2$ as explained in \se\ref{se:method}, 
the sum-integrals in \eqs\nr{Ja}, \nr{Jb} can be expressed as
\ba
  \mathcal{J}_\rmi{a} & = & 
% \Tint{Q} \frac{P^2}{Q^2}
 P^2 \int_{\vec{q}} \frac{\nB{}(q)}{q}
 \;, \la{Ja_res} \\ 
 \mathcal{J}_\rmi{b} & = & 
% \Tint{Q} \frac{P^4}{Q^2(Q-P)^2}
 \frac{P^{4-2\epsilon}}{(4\pi)^{2-\epsilon}}
 \frac{\Gamma(1+\epsilon) \Gamma^2(1-\epsilon)}{\Gamma(1-2\epsilon)}
 \frac{1}{\epsilon(1-2\epsilon)}
 \nn & & \; 
 + \, 2 P^2 \int_{\vec{q}} \frac{\nB{}(q)}{q}
 + \frac{8}{P^2}\biggl( \frac{p^2}{3-2\epsilon} - p_n^2 \biggr)
 \int_{\vec{q}} q\, \nB{}(q)
 + \rmO\Bigl(\frac{1}{P^2}\Bigr)
 \;. % \nn 
\ea
Introducing the shorthands
\newcommand{\s}{\mathcal{S}}
\ba
 \s_1 & \equiv & \frac{P^{4-4\epsilon}}{(4\pi)^{4-2\epsilon}}
 \frac{\Gamma^2(1+\epsilon) \Gamma^4(1-\epsilon)}{\Gamma^2(1-2\epsilon)}
 \;, \la{s1_def} \\ 
 \s_2 & \equiv & \frac{\Gamma(1+2\epsilon)\Gamma^2(1-2\epsilon)}
 {\Gamma(1-3\epsilon)\Gamma^2(1+\epsilon)\Gamma(1-\epsilon)}
 \;, \\ 
 \s_3 & \equiv & \frac{P^{2-2\epsilon}}{(4\pi)^{2-\epsilon}} 
 \frac{\Gamma(1+\epsilon) \Gamma^2(1-\epsilon)}{\Gamma(1-2\epsilon)}
 \int_{\vec{q}} \frac{\nB{}(q)}{q}
 \;, \\ 
 \s_4 & \equiv & \frac{(1-2\epsilon)P^2}{2} \int_{\vec{q},\vec{r}}
 \frac{\nB{}(q)}{q} \frac{\nB{}(r)}{r^3} 
 \;, \\ 
 \s_5 & \equiv & 
 \frac{P^{-2\epsilon}}{(4\pi)^{2-\epsilon}} 
 \frac{\Gamma(1+\epsilon) \Gamma^2(1-\epsilon)}{\Gamma(1-2\epsilon)}
 \frac{1}{P^2}\biggl( \frac{p^2}{3-2\epsilon} - p_n^2 \biggr)
 \int_{\vec{q}} q\, \nB{}(q)
 \;, \\ 
 \s_6 & \equiv & 
 \int_{\vec{q},\vec{r}}
 \frac{\nB{}(q)}{q} \frac{\nB{}(r)}{r}
 \;, \la{s6_def}
\ea 
we can write the expansions of
the next-to-leading order sum-integrals as
\ba
 \mathcal{I}_\rmi{a} & = & 
% \Tint{Q,R} \frac{1}{Q^2R^2}
 \s_6
 \;, \la{Ia_res} \\ 
 \mathcal{I}_\rmi{b} & = & 
% \Tint{Q,R} \frac{P^2}{Q^2R^2(R-P)^2}
 \frac{\s_3}{\epsilon(1-2\epsilon)} + 2 \s_6
 + \rmO\Bigl(\frac{1}{P^2}\Bigr)
 \;, \\ 
 \mathcal{I}_\rmi{c} & = & 
 \s_4
% \Tint{Q,R} \frac{P^2}{Q^2R^4}
 \;, \\ 
 \mathcal{I}_\rmi{d} & = & 
% \Tint{Q,R} \frac{P^4}{Q^2R^4(R-P)^2}
 -\frac{\s_3}{\epsilon} + \s_4 +  
% \frac{2}{P^2}\Bigl[p^2 - (1-2\epsilon)p_n^2 \Bigr] \s_6
 \biggl[ \frac{2}{2-\epsilon}
 + \frac{2(1-\epsilon)(3-2\epsilon)}{(2-\epsilon)P^2}
 \biggl( \frac{p^2}{3-2\epsilon} - p_n^2 \biggr) \biggr] \s_6
 + \rmO\Bigl(\frac{1}{P^2}\Bigr)
 \;, \\ 
 \mathcal{I}_\rmi{e} & = & 
% \Tint{Q,R} \frac{P^2}{Q^2R^2(Q-R)^2}
 0
 \;, \\ 
 \mathcal{I}_\rmi{f} & = & 
% \Tint{Q,R} \frac{P^2}{Q^2(Q-R)^2(R-P)^2}
 -\frac{\s_1\s_2}{2\epsilon(1-2\epsilon)(1-3\epsilon)(2-3\epsilon)}
 + \frac{3 \s_3}{\epsilon(1-2\epsilon)}
 + \frac{6(1+\epsilon)\s_5}{1-2\epsilon}
 + 3 \s_6
 + \rmO\Bigl(\frac{1}{P^2}\Bigr)
 \;, \hspace*{5mm} \\ 
 \mathcal{I}_\rmi{g} & = & 
% \Tint{Q,R} \frac{P^4}{Q^2(Q-P)^2R^2(R-P)^2}
 \frac{\s_1}{\epsilon^2(1-2\epsilon)^2}
 + \frac{4\s_3}{\epsilon(1-2\epsilon)}
 + \frac{16 \s_5}{\epsilon(1-2\epsilon)}
 + 4 \s_6
 + \rmO\Bigl(\frac{1}{P^2}\Bigr)
 \;,  \\ 
 \mathcal{I}_\rmi{h} & = & 
% \Tint{Q,R} \frac{P^4}{Q^2R^2(Q-R)^2(R-P)^2}
 \frac{\s_1\s_2}{2\epsilon^2(1-2\epsilon)(1-3\epsilon)}
 - \frac{(1-4\epsilon)\s_3}{\epsilon(1-2\epsilon)}
 + \frac{2(1+\epsilon)(2+\epsilon)(1+4\epsilon)\s_5}{3\epsilon(1-2\epsilon)}
 \nn & & 
 \; + \;  
% \biggl\{ 1 + 
%  \frac{2}{\epsilon P^2}
% \biggl[ \frac{p^2}{3-2\epsilon} - (1-2\epsilon) p_n^2 \biggr]
% \biggr\}
 \biggl[ 
 \frac{4-\epsilon}{2-\epsilon} + 
 \frac{4(1-\epsilon)^2}{\epsilon(2-\epsilon)P^2}
 \biggl( \frac{p^2}{3-2\epsilon} - p_n^2 \biggr)
 \biggl]
 \s_6
 + \rmO\Bigl(\frac{1}{P^2}\Bigr)
 \;, \ea \ba
%
% \mathcal{I}_\rmi{i} & = & 
% \mathcal{I}_\rmi{a} + \mathcal{I}_\rmi{e} - \mathcal{I}_\rmi{f}
% + \mathcal{I}_\rmi{i'}
% \;, \\ 
%
 \mathcal{I}_\rmi{i'} & = & 
% \Tint{Q,R} \frac{4(Q\cdot P)^2}{Q^2R^2(Q-R)^2(R-P)^2}
 \frac{\s_1\s_2}{3\epsilon^2(1-3\epsilon)(2-3\epsilon)}
 + \frac{2(1+\epsilon)\s_3}{\epsilon(1-2\epsilon)}
 - \frac{4(9-35\epsilon+16\epsilon^2-\epsilon^3+2\epsilon^4)\s_5}
   {3\epsilon(1-2\epsilon)(3-2\epsilon)}
 \nn & & 
 \; + \;  
% \biggl\{ 4 + 
%  \frac{2}{\epsilon P^2}
% \biggl[ \frac{p^2}{3-2\epsilon} - (1-2\epsilon) p_n^2 \biggr]
% \biggr\} 
 \biggl[ 
 \frac{2(5-2\epsilon)}{2-\epsilon} + 
 \frac{4(1-\epsilon)^2}{\epsilon(2-\epsilon)P^2}
 \biggl( \frac{p^2}{3-2\epsilon} - p_n^2 \biggr)
 \biggl]
 \s_6 
 + \rmO\Bigl(\frac{1}{P^2}\Bigr)
 \;, \\ 
 \mathcal{I}_\rmi{j} & = & 
% \Tint{Q,R} \frac{P^6}{Q^2R^2(Q-R)^2(Q-P)^2(R-P)^2}
 \frac{\s_1(1-\s_2)}{\epsilon^3(1-2\epsilon)}
 - \frac{2(3+\epsilon)\s_3}{\epsilon}
 - \frac{2(5+\epsilon)(10+5\epsilon+\epsilon^2)\s_5}
   {3\epsilon}
 \nn & & 
 \; + \; 
% \biggl\{ -2 + 
%  \frac{10}{\epsilon P^2}
% \biggl[ \frac{p^2}{3-2\epsilon} - (1-2\epsilon) p_n^2 \biggr]
% \biggr\}
 \biggl[ 
 \frac{2(3+\epsilon)}{2-\epsilon} + 
 \frac{20(1-\epsilon)^2}{\epsilon(2-\epsilon)P^2}
 \biggl( \frac{p^2}{3-2\epsilon} - p_n^2 \biggr)
 \biggl]
 \s_6 
 + \rmO\Bigl(\frac{1}{P^2}\Bigr)
 \;. \la{Ij_res}
\ea
Both structures proportional to $P^2$, namely $\s_3$ appearing
in almost every master sum-integral as well as $\s_4$ appearing
in $\mathcal{I}_\rmi{c}$ and $\mathcal{I}_\rmi{d}$,  
disappear from the final result for any $\epsilon$. 
A useful relation allowing to change the basis in the terms 
proportional to $\s_6$ is
\be
 \frac{2}{\epsilon P^2}
 \biggl[ \frac{p^2}{3-2\epsilon} - (1-2\epsilon) p_n^2 \biggr]
 = \frac{2}{2-\epsilon} + 
 \frac{4(1-\epsilon)^2}{\epsilon(2-\epsilon)P^2}
 \biggl[ \frac{p^2}{3-2\epsilon} - p_n^2 \biggr]
 \;.
\ee

%%%%%%%%%%%%%%%%%%%%%%%%%%%%%% SECTION %%%%%%%%%%%%%%%%%%%%%%%%%%%%%%%%%
%
\section{Basic thermodynamic functions}

We recall here perturbative expressions for 
a number of thermodynamic potentials needed in our analysis.
The results are needed up to order $\rmO(g^6)$ in some cases and 
can be extracted from the explicit $\rmO(g^6)$ results in ref.~\cite{gsixg}; 
however, they could also be deduced through renormalization 
scale independence arguments already from the classic $\rmO(g^2)$ 
results for the thermodynamic pressure given in refs.~\cite{es,chin}.

As explained in the main body of the text, 
an essential role in the present study is played by
the energy-momentum tensor of the thermalized system, which 
in the plasma rest frame takes the form $\mathop{\rm diag}(e,-p,-p,-p)$.
We separate this into a traceless part, whose $00$-component reads
\be
 e - \fr14 (e-3p) = \fr34 (e+p)
 \;, 
\ee  
as well as a trace part, 
$
 e- 3p
$.
At next-to-leading order, the combination appearing 
in the traceless part can be written as 
\be
 (e+p)(T) = \frac{8 d_A}{3} 
 \biggl[
   \int_{\vec{q}} q\, \nB{}(q) - \frac{3g^2\Nc}{2}
   \int_{\vec{q},\vec{r}}
   \frac{\nB{}(q)}{q} \frac{\nB{}(r)}{r}
 \biggr]
 \;, \la{eT}
\ee
with $d_A = \Nc^2-1$, 
whereas the leading-order expression for the trace anomaly reads
\be
 (e - 3 p)(T)
 = 2 d_A g^4 b_0 \Nc 
   \int_{\vec{q},\vec{r}}
   \frac{\nB{}(q)}{q} \frac{\nB{}(r)}{r}
 \;, \la{em3pT}
\ee 
with $b_0$ as defined in \eq\nr{b0}.
Finally, the temperature dependence of the trace anomaly reads
\be
 T^5 \frac{{\rm d}}{{\rm d}T} \biggl(\frac{e-3p}{T^4}\biggr)
 = 
 -8 d_A g^6 b_0^2 \Nc 
   \int_{\vec{q},\vec{r}}
   \frac{\nB{}(q)}{q} \frac{\nB{}(r)}{r}
 \;. \la{Tem3pT}
\ee
In the expressions above, $\nB{}$ denotes the Bose distribution, 
$\nB{}(q) \equiv 1/(e^{\beta q} - 1)$, and the integrals can be carried out,  
\be
 \int_{\vec{q}} q\, \nB{}(q) = \frac{\pi^2 T^4}{30} + \rmO(\epsilon)
 \;, \quad 
 \int_{\vec{q}}  \frac{\nB{}(q)}{q} = \frac{T^2}{12} + \rmO(\epsilon)
 \;, \la{intrep}
\ee
but it is convenient 
to be able to recognize the integral representations as well. 

%%%%%%%%%%%%%%%%%%%%%%%%%%%%%% SECTION %%%%%%%%%%%%%%%%%%%%%%%%%%%%%%%%%
%
\section{On fermionic effects}

We add here the contribution of $\Nf$ massless ``sea'' quarks 
to the previous results for the trace anomaly and topological 
charge density correlators. 
By sea quarks we refer to effects originating from the fermionic
contribution to the gluon self-energy; ``valence'' quarks
would refer to the quark part of 
the energy-momentum tensor operator. It is often said that
the quark contribution to the trace anomaly operator, which is proportional
to $\bar\psi \gamma_\mu D_\mu \psi$ or
$\bar\psi \gamma_\mu \overleftrightarrow{D}_\mu \psi $, 
vanishes in the chiral limit
thanks to the equations of motion; in topological susceptibility
there should be no valence quark contribution to start with. 
Below we find practically identical expressions for the two channels, 
however also structures in the Lorentz-symmetry violating part
of the result which do not fit the form expected from the 
Operator Product Expansion. This could indicate that there is, 
after all, some mixing taking place and a non-zero valence quark 
contribution to be added at non-zero $T$; however, given 
the fair amount of work involved,  we have not 
carried out a systematic investigation to clear up the issue. 

With these reservations, 
the fermionic contributions to the correlators in \eq\nr{GP} read
\ba
 && \hspace*{-1cm} 
 \frac{\delta\tilde G_\theta(P)}{-4 d_A c_\theta^2 \gB^6 \Nf}  
 \nn & = & 
 4 \biggl[ \widetilde{\mathcal{I}}_\rmi{a}
 + \widetilde{\mathcal{I}}_\rmi{i} \biggr]
 + 2(D-1) \biggl[ -4 \widetilde{\mathcal{I}}_\rmi{a}
  +  \overline{\mathcal{I}}_\rmi{a} - \widetilde{\mathcal{I}}_\rmi{e} \biggr]
 \nn & & \quad + \, 
 (D-2) 
 \biggl[ 
   \overline{\mathcal{I}}_\rmi{e} + 2\widetilde{\mathcal{I}}_\rmi{c}
    - 2\widetilde{\mathcal{I}}_\rmi{d}
 \biggr] +
 (D-4) 
 \biggl[ 
   4 \widetilde{\mathcal{I}}_\rmi{b} - 2\widetilde{\mathcal{I}}_\rmi{f}
    +\overline{\mathcal{I}}_\rmi{f} + \widetilde{\mathcal{I}}_\rmi{h}
 \biggr] % \biggr\} % + \rmO(\gB^4)
 \;, \la{Gtheta_bare_f} \\
%%%%%%%%%%%%%%%%%%%%%%%%%%%%%%%%%%%%%%
%%%%%%%%%%%%%%%%%%%%%%%%%%%%%%%%%%%%%% 
 && \hspace*{-1cm} 
 \frac{\delta \tilde G_\chi(P)}{16 d_A c_\chi^2 \gB^6 (D-3) \Nf} 
 \nn & = & 
 4 \biggl[ \overline{\mathcal{I}}_\rmi{a} - \widetilde{\mathcal{I}}_\rmi{a}
 + \widetilde{\mathcal{I}}_\rmi{i} \biggr]
 - 2(D-1) \biggl[ 
    \overline{\mathcal{I}}_\rmi{a} + \widetilde{\mathcal{I}}_\rmi{e} \biggr]
 \nn & & \quad + \, 
 (D-2) 
 \biggl[ 
   \overline{\mathcal{I}}_\rmi{e} + 2\widetilde{\mathcal{I}}_\rmi{c}
    - 2\widetilde{\mathcal{I}}_\rmi{d}
 \biggr] +
 (D-4) 
 \biggl[ 
   4 \widetilde{\mathcal{I}}_\rmi{b} - 2\widetilde{\mathcal{I}}_\rmi{f}
    +\overline{\mathcal{I}}_\rmi{f} + \widetilde{\mathcal{I}}_\rmi{h}
 \biggr] % \biggr\} % + \rmO(\gB^4)
 \;. \la{Gchi_bare_f}
\ea 
Here the following new master sum-integrals have been introduced: 
\ba
 \widetilde{\mathcal{I}}_\rmi{a} & \equiv & 
 \Tint{\{Q\},R} \frac{1}{Q^2R^2}
 \;, \\ 
 \overline{\mathcal{I}}_\rmi{a} & \equiv & 
 \Tint{\{Q\},R} \frac{1}{Q^2(R-Q)^2}
 \;, \\ 
 \widetilde{\mathcal{I}}_\rmi{b} & \equiv & 
 \Tint{\{Q\},R} \frac{P^2}{Q^2R^2(R-P)^2}
 \;, \\ 
 \widetilde{\mathcal{I}}_\rmi{c} & \equiv & 
 \Tint{\{Q\},R} \frac{P^2}{Q^2R^4}
 \;, \\ 
 \widetilde{\mathcal{I}}_\rmi{d} & \equiv & 
 \Tint{\{Q\},R} \frac{P^4}{Q^2R^4(R-P)^2}
 \;, \\ 
 \widetilde{\mathcal{I}}_\rmi{e} & \equiv & 
 \Tint{\{Q\},R} \frac{P^2}{Q^2R^2(Q-R)^2}
 \;, \\ 
 \overline{\mathcal{I}}_\rmi{e} & \equiv & 
 \Tint{\{Q\},R} \frac{(R-P)^2}{Q^2R^2(Q-R)^2}
 \; = \;
 \overline{\mathcal{I}}_\rmi{a} + \widetilde{\mathcal{I}}_\rmi{e}
 \;, \\ 
 \widetilde{\mathcal{I}}_\rmi{f} & \equiv & 
 \Tint{\{Q\},R} \frac{P^2}{Q^2(Q-R)^2(R-P)^2}
 \;, \\ 
 \overline{\mathcal{I}}_\rmi{f} & \equiv & 
 \Tint{\{Q\},R} \frac{R^2}{Q^2(Q-R)^2(R-P)^2}
 \; = \; 
 \overline{\mathcal{I}}_\rmi{a} 
 - \widetilde{\mathcal{I}}_\rmi{f}
 + \overline{\mathcal{I}}_\rmi{f'}
 \;, \\ 
 \overline{\mathcal{I}}_\rmi{f'} & \equiv & 
 \Tint{\{Q\},R} \frac{2R\cdot P}{Q^2(Q-R)^2(R-P)^2}
 \;, \\ 
 \widetilde{\mathcal{I}}_\rmi{h} & \equiv & 
 \Tint{\{Q\},R} \frac{P^4}{Q^2R^2(Q-R)^2(R-P)^2}
 \;, \\ 
 \widetilde{\mathcal{I}}_\rmi{i} & \equiv & 
 \Tint{\{Q\},R} \frac{(Q-P)^4}{Q^2R^2(Q-R)^2(R-P)^2}
 \; = \; 
   \widetilde{\mathcal{I}}_\rmi{a}
 + \widetilde{\mathcal{I}}_\rmi{e} 
 - \widetilde{\mathcal{I}}_\rmi{f}
 + \widetilde{\mathcal{I}}_\rmi{i'}
 \;, \\ 
 \widetilde{\mathcal{I}}_\rmi{i'} & \equiv & 
 \Tint{\{Q\},R} \frac{4(Q\cdot P)^2}{Q^2R^2(Q-R)^2(R-P)^2}
 \;. \la{Iip_f}
\ea
The sum-integral denoted by $\Tinti{\{Q\}}$ goes over fermionic
Matsubara momenta. As indicated this set is somewhat overcomplete
(at least in the absence of a chemical potential, as we have
assumed to be the case throughout). 

We note that, like in the bosonic case, most of the master
sum-integrals vanish in the limit $P\to 0$, relevant for the 
sum rule (cf.\ \eqs\nr{sum_theta}, \nr{sum_chi}). Non-zero
contributions arise only from $\widetilde{\mathcal{I}}_\rmi{a}$ and
$\overline{\mathcal{I}}_\rmi{a}$. In $\delta \tilde G_\chi(0)$
these cancel; for $\delta \tilde G_\theta(0)$ we obtain 
(inserting also $c_\theta$ from \eq\nr{c_theta})
\be
  \delta \tilde G_\theta(0) =   
  8 d_A g^6 b_0^2 \Nf  
  \, (2 \widetilde{\mathcal{I}}_a - \overline{\mathcal{I}}_a)
% \Tint{\{Q\},R}
% \biggl[  \frac{2}{Q^2R^2} - \frac{1}{Q^2(R-Q)^2}  
% \biggr]
 = -  \frac{5 d_A g^6 b_0^2 \Nf T^4}{72}
 \;, \la{sum_theta_f}
\ee
where we made use of 
$\Tinti{\{Q\}} 1/Q^2 = -T^2/24$ and 
$\Tinti{R} 1/R^2 = T^2/12$. 
% and 
% $
%  b_0 = \frac{11\Nc - 2\Nf}{3(4\pi)^2}
% $.
This result agrees with the leading fermionic contribution to 
$
 T^5 {\rm d} [ (e-3p)/T^4 ] / {\rm d}T
$, 
and therefore conforms with the sum rule 
of ref.~\cite{ellis} at $\rmO(g^6)$.

The large-momentum expansions of the master sum-integrals can be 
worked out like in the bosonic case (cf.\ \se\ref{se:method}).
Supplementing the structures in \eqs\nr{s1_def}--\nr{s6_def} with
\ba
 \widetilde{\s}_3 & \equiv & - \frac{P^{2-2\epsilon}}{(4\pi)^{2-\epsilon}} 
 \frac{\Gamma(1+\epsilon) \Gamma^2(1-\epsilon)}{\Gamma(1-2\epsilon)}
 \int_{\vec{q}} \frac{\nF{}(q)}{q}
 \;, \\ 
 \widetilde{\s}_4 & \equiv & - \frac{(1-2\epsilon)P^2}{2} \int_{\vec{q},\vec{r}}
 \frac{\nF{}(q)}{q} \frac{\nB{}(r)}{r^3} 
 \;, \\ 
 \widetilde{\s}_5 & \equiv & 
 - \frac{P^{-2\epsilon}}{(4\pi)^{2-\epsilon}} 
 \frac{\Gamma(1+\epsilon) \Gamma^2(1-\epsilon)}{\Gamma(1-2\epsilon)}
 \frac{1}{P^2}\biggl( \frac{p^2}{3-2\epsilon} - p_n^2 \biggr)
 \int_{\vec{q}} q\, \nF{}(q)
 \;, \\ 
 \widetilde{\s}_6 & \equiv & 
 - \int_{\vec{q},\vec{r}}
 \frac{\nF{}(q)}{q} \frac{\nB{}(r)}{r}
 \;, \\ 
 \overline{\s}_6 & \equiv & 
 \int_{\vec{q},\vec{r}}
 \frac{\nF{}(q)}{q} \frac{\nF{}(r)}{r}
 \;, \la{s6_def_f}
\ea 
where $\nF{}$ denotes the Fermi distribution, 
$\nF{}(q)\equiv 1/(e^{\beta q} + 1)$, we obtain the expansions
\ba
 \widetilde{\mathcal{I}}_\rmi{a} & = & 
% \Tint{Q,R} \frac{1}{Q^2R^2}
 \widetilde{\s}_6
 \;, \\ 
 \overline{\mathcal{I}}_\rmi{a} & = & 
% \Tint{Q,R} \frac{1}{Q^2(R-Q)^2}
 \overline{\s}_6
 \;, \\ 
 \widetilde{\mathcal{I}}_\rmi{b} & = & 
% \Tint{Q,R} \frac{P^2}{Q^2R^2(R-P)^2}
 \frac{\widetilde{\s}_3}{\epsilon(1-2\epsilon)} + 2 \widetilde{\s}_6
 + \rmO\Bigl(\frac{1}{P^2}\Bigr)
 \;, \\ 
 \widetilde{\mathcal{I}}_\rmi{c} & = & 
 \widetilde{\s}_4
% \Tint{Q,R} \frac{P^2}{Q^2R^4}
 \;, \\
 \widetilde{\mathcal{I}}_\rmi{d} & = & 
% \Tint{Q,R} \frac{P^4}{Q^2R^4(R-P)^2}
 -\frac{\widetilde{\s}_3}{\epsilon} + \widetilde{\s}_4 +  
 \biggl[ \frac{2}{2-\epsilon}
 + \frac{2(1-\epsilon)(3-2\epsilon)}{(2-\epsilon)P^2}
 \biggl( \frac{p^2}{3-2\epsilon} - p_n^2 \biggr) \biggr] \widetilde{\s}_6
 + \rmO\Bigl(\frac{1}{P^2}\Bigr)
 \;, \\ 
 \widetilde{\mathcal{I}}_\rmi{e} & = & 
% \Tint{Q,R} \frac{P^2}{Q^2R^2(Q-R)^2}
 0
 \;, \\
 \widetilde{\mathcal{I}}_\rmi{f} & = & \!\!\! 
% \Tint{Q,R} \frac{P^2}{Q^2(Q-R)^2(R-P)^2}
 -\frac{\s_1\s_2}{2\epsilon(1-2\epsilon)(1-3\epsilon)(2-3\epsilon)}
 + \frac{\s_3 + 2 \widetilde{\s}_3}{\epsilon(1-2\epsilon)}
 \nn & &  \; 
 + \, \frac{2(1+\epsilon)(\s_5 + 2 \widetilde{\s}_5)}{1-2\epsilon}
 + \overline{\s}_6 + 2 \widetilde{\s}_6
 + \rmO\Bigl(\frac{1}{P^2}\Bigr)
 \;, \hspace*{6mm} \ea \ba 
 \overline{\mathcal{I}}_\rmi{f'} \!\! & = & 
% \Tint{Q,R} \frac{P^2}{Q^2(Q-R)^2(R-P)^2}
 -\frac{2\s_1\s_2}{3\epsilon(1-2\epsilon)(1-3\epsilon)(2-3\epsilon)}
 + \frac{2(\s_3 + \widetilde{\s}_3)}{\epsilon(1-2\epsilon)}
 \nn & &  \; 
 + \, \frac{4\epsilon\s_5 + 4 (2+\epsilon) \widetilde{\s}_5}{1-2\epsilon}
 + 4 \widetilde{\s}_6
 + \rmO\Bigl(\frac{1}{P^2}\Bigr)
 \;, \\
 \widetilde{\mathcal{I}}_\rmi{h} & = & 
% \Tint{Q,R} \frac{P^4}{Q^2R^2(Q-R)^2(R-P)^2}
 \frac{\s_1\s_2}{2\epsilon^2(1-2\epsilon)(1-3\epsilon)}
 + \frac{\s_3-2 (1-2\epsilon)\widetilde{\s}_3}{\epsilon(1-2\epsilon)}
 + \frac{2(1+\epsilon)(2+\epsilon)
 [3 \s_5 - 2 (1-2\epsilon)\widetilde{\s}_5]}{3\epsilon(1-2\epsilon)}
 \nn & & 
 \; + \;  
% \biggl\{ 1 + 
%  \frac{2}{\epsilon P^2}
% \biggl[ \frac{p^2}{3-2\epsilon} - (1-2\epsilon) p_n^2 \biggr]
% \biggr\}
  2 \widetilde{\s}_6 + 
 \biggl[ 
 \frac{\epsilon}{2-\epsilon} + 
 \frac{4(1-\epsilon)^2}{\epsilon(2-\epsilon)P^2}
 \biggl( \frac{p^2}{3-2\epsilon} - p_n^2 \biggr)
 \biggl]
 \overline{\s}_6
 + \rmO\Bigl(\frac{1}{P^2}\Bigr)
 \;, \\  
 \widetilde{\mathcal{I}}_\rmi{i'} & = & 
% \Tint{Q,R} \frac{4(Q\cdot P)^2}{Q^2R^2(Q-R)^2(R-P)^2}
 \frac{\s_1\s_2}{3\epsilon^2(1-3\epsilon)(2-3\epsilon)}
 + \frac{\s_3 + (1+2\epsilon)\widetilde{\s}_3}{\epsilon(1-2\epsilon)}
 +  \frac{2(2+2\epsilon+\epsilon^2-2\epsilon^3)\s_5}
   {\epsilon(1-2\epsilon)(3-2\epsilon)}
 - \frac{2(8 + \epsilon^2)\widetilde{\s}_5}
   {3\epsilon}
 \nn & & 
 \; + \;  
% \biggl\{ 4 + 
%  \frac{2}{\epsilon P^2}
% \biggl[ \frac{p^2}{3-2\epsilon} - (1-2\epsilon) p_n^2 \biggr]
% \biggr\} 
 \biggl[ 
 \frac{2(5-2\epsilon)}{2-\epsilon} + 
 \frac{4(1-\epsilon)^2}{\epsilon(2-\epsilon)P^2}
 \biggl( \frac{p^2}{3-2\epsilon} - p_n^2 \biggr)
 \biggl]
 \widetilde{\s}_6 
 + \rmO\Bigl(\frac{1}{P^2}\Bigr)
 \;. \la{Ij_res_f}
\ea
All structures proportional to $P^2$, namely 
$\s_3, \widetilde{\s}_3$ and $\widetilde{\s}_4$, 
cancel in the final results for any $\epsilon$. 

Inserting these expansions into \eqs\nr{Gtheta_bare_f}, \nr{Gchi_bare_f}
and re-expressing the bare gauge coupling in terms of the renormalized
one according to \eq\nr{gB}, 
we obtain results analogous to \eqs\nr{theta_P}, \nr{chi_P}. Alas, 
many different thermal distributions appear and it is not easy 
to express the result in a concise way. We choose
rather to insert explicit values, 
\ba
 \int_{\vec{q}} q\, \nB{}(q) & = &  
  \mu^{-2\epsilon} \; \frac{\pi^2 T^4 }{30}
 \biggl\{ 1 + 2\epsilon 
 \biggl[ 
   \ln\frac{\bmu}{4\pi T} + 1 + \frac{\zeta'(-3)}{\zeta(-3)}
 \biggr] 
 + \rmO(\epsilon^2) \biggr\}
 \;, \\ 
 \int_{\vec{q}}  \frac{\nB{}(q)}{q} & = &   \mu^{-2\epsilon} \; \frac{T^2 }{12} 
 \biggl\{ 1 + 2\epsilon 
 \biggl[ 
   \ln\frac{\bmu}{4\pi T} + 1 + \frac{\zeta'(-1)}{\zeta(-1)}
 \biggr] 
 + \rmO(\epsilon^2) \biggr\}
 \;, \la{eps_intrep1} \\ 
 \int_{\vec{q}} q\, \nF{}(q) & = &  \mu^{-2\epsilon}\; \frac{7\pi^2 T^4 }{240}
 \biggl\{ 1 + 2\epsilon 
 \biggl[ 
   \ln\frac{\bmu}{4\pi T} + 1 + \frac{\zeta'(-3)}{\zeta(-3)} - \frac{\ln 2}{7}
 \biggr] 
 + \rmO(\epsilon^2) \biggr\}
 \;, \\ 
 \int_{\vec{q}}  \frac{\nF{}(q)}{q} & = &  \mu^{-2\epsilon}\; \frac{T^2 }{24}
 \biggl\{ 1 + 2\epsilon 
 \biggl[ 
   \ln\frac{\bmu}{4\pi T} + 1 + \frac{\zeta'(-1)}{\zeta(-1)} - \ln 2
 \biggr] 
 + \rmO(\epsilon^2) \biggr\}
 \;. \la{eps_intrep2}
\ea
Thereby the fermionic contributions to \eqs\nr{theta_P}, \nr{chi_P} become
\ba
 && \hspace*{-1cm} 
 \frac{\delta\tilde G_\theta(P)}
 {4 d_A c_\theta^2 g^4 \mu^{2\epsilon}}   \nn
 & = & 
 \frac{P^4}{(4\pi)^2}
 \biggl\{
 \biggl(\frac{\bmu}{P} \biggr)^{2\epsilon}
 \biggl[ \frac{1}{\epsilon} + 1 +... \biggr]
 \times \frac{g^2 \Nf}{(4\pi)^2} \frac{4}{3\epsilon}
 -
 \frac{g^2 \Nf}{(4\pi)^2} 
 \biggl(\frac{\bmu}{P} \biggr)^{4\epsilon}
 \biggl[
  \frac{2}{3\epsilon^2} + \frac{3}{\epsilon} +... 
 \biggr]
 \biggr\}
 \nn 
 & + & 
 \frac{g^2\Nf T^4}{45 P^2} \biggl( \frac{p^2}{3} - p_n^2 \biggr)
 \biggl[
  \fr52 \ln\frac{\bmu}{4\pi T} 
  - \fr92 \ln \frac{\bmu}{P}
  -2\ln 2  -\frac{151}{48}
  + 5 \frac{\zeta'(-1)}{\zeta(-1)}
  - \fr52 \frac{\zeta'(-3)}{\zeta(-3)}
 \biggr]
 \nn 
 & - & \frac{5 g^2 \Nf T^4}{144} 
  + \rmO\Bigl(g^4,\frac{1}{P^2}\Bigr)
 \;, \la{theta_P_f} \ea \ba 
%%%%%
%%%%%
  && \hspace*{-1cm} 
 \frac{\delta\tilde G_\chi(P)}
 {-16 d_A c_\chi^2 g^4  \mu^{2\epsilon}}   \nn 
 & = & 
 \frac{P^4}{(4\pi)^2}
 \biggl\{
 \biggl(\frac{\bmu}{P} \biggr)^{2\epsilon}
 \biggl[ \frac{1}{\epsilon} - 1 +... \biggr]
 \times \frac{g^2 \Nf}{(4\pi)^2} \frac{4}{3\epsilon} 
 -
 \frac{g^2 \Nf}{(4\pi)^2} 
 \biggl(\frac{\bmu}{P} \biggr)^{4\epsilon}
 \biggl[
  \frac{2}{3\epsilon^2} + \frac{5}{3\epsilon} +... 
 \biggr]
 \biggr\}
 \nn 
 & + & 
 \frac{g^2\Nf T^4}{45 P^2} \biggl( \frac{p^2}{3} - p_n^2 \biggr)
 \biggl[
  \fr52 \ln \frac{\bmu}{4\pi T} 
  - \fr92 \ln \frac{\bmu}{P}
  -2\ln 2  -\frac{151}{48}
  + 5 \frac{\zeta'(-1)}{\zeta(-1)}
  - \fr52 \frac{\zeta'(-3)}{\zeta(-3)}
 \biggr]
 \nn 
 & + & \frac{5 g^2 \Nf T^4}{144} 
  + \rmO\Bigl(g^4,\frac{1}{P^2}\Bigr)
 \;. \la{chi_P_f} 
\ea

Inspecting the results, 
the terms on the last rows of \eqs\nr{theta_P_f}, \nr{chi_P_f} are easy 
to understand: the leading-order quark contribution to the trace anomaly is 
\be
 \delta\biggl( \frac{e-3p}{T^4} \biggr)
 = 
 \frac{5 d_A g^4  b_0 \Nf}{288}
 \;,
\ee
and these terms amount to $\mp 2 \delta(e-3p)/d_A g^2 b_0 $, 
in perfect accordance with the bosonic results
of \eqs\nr{D_theta_P}, \nr{D_chi_P}. In contrast, 
the terms on the second rows of \eqs\nr{theta_P_f}, \nr{chi_P_f}
fit no simple pattern. In principle we might expect a fermionic
contribution to the Wilson coefficient multiplying the leading
bosonic $e+p$, {\it viz.}\ $8 d_A \pi^2 T^4/90$, as well as 
direct fermionic effects to $e+p$,  
\be
 \delta\biggl( \frac{e+p}{T^4} \biggr)
 = 
 \frac{7\Nc\Nf \pi^2}{45} - \frac{5d_A g^2 \Nf}{144}
 \;, \la{ep_f}
\ee
the former multiplied by the next-to-leading order Wilson 
coefficient and the latter by the leading order one. 
However, there is no way to reproduce the {\em leading} fermionic 
contribution to $e+p$ (the first term in \eq\nr{ep_f}) from the 
effects in \eqs\nr{theta_P_f}, \nr{chi_P_f} 
(powers of $g^2$ and/or group theory factors do not match) 
and, conversely, there is 
no way to understand the appearance of  the temperature-dependent
logarithms in \eqs\nr{theta_P_f}, \nr{chi_P_f} in terms 
of the Operator Product Expansion contributions. Something 
is clearly missing and, as mentioned at the beginning, 
one possibility could be an unaccounted
mixing with fermionic operators. Let us stress that 
the problem only appears in the contributions proportional to $e+p$, 
which vanish at zero temperature due to Lorentz symmetry
(cf.\ the last lines of \se\ref{se:mom}).  

\newpage

%%%%%%%%%%%%%%%%%%%%%%%%%%%%%%%%%%%%%%%%%%%%%%%%%%%%%%%%%%%%%%%%%%%%%%%%%%%
%


\begin{thebibliography}{99}

\bibitem{ay}
  P.B.~Arnold and L.G.~Yaffe,
  {\it Non-Abelian Debye screening length beyond leading order,}
  Phys.\ Rev.\  D {52} (1995) 7208
  [hep-ph/9508280].
  %%CITATION = PHRVA,D52,7208;%%

\bibitem{ms}
  G.D.~Moore and O.~Saremi,
  {\it Bulk viscosity and spectral functions in QCD,}
  JHEP {09} (2008) 015
  [0805.4201].
  %%CITATION = JHEPA,0809,015;%%

\bibitem{linde}
  A.D.~Linde,
  {\it Infrared problem in thermodynamics of the Yang-Mills gas,}
  Phys.\ Lett.\ {B 96} (1980) 289.
  %%CITATION = PHLTA,B96,289;%%

\bibitem{gpy}
  D.J.~Gross, R.D.~Pisarski and L.G.~Yaffe,
  {\it QCD and instantons at finite temperature,}
  Rev.\ Mod.\ Phys.\ {53} (1981) 43.
  %%CITATION = RMPHA,53,43;%%

\bibitem{chm}
  S.~Caron-Huot and G.D.~Moore,
  {\it Heavy quark diffusion in QCD and $\mathcal{N}=4$ SYM 
  at next-to-leading order,}
  JHEP {02} (2008) 081
  [0801.2173].
  %%CITATION = JHEPA,0802,081;%%

\bibitem{lv}
  M.~Laine and M.~Veps\"al\"ainen,
  {\it On the smallest screening masses in hot QCD,}
  JHEP {09} (2009) 023
  [0906.4450].
  %%CITATION = JHEPA,0909,023;%%


\bibitem{singlet}
  Y.~Burnier, M.~Laine and M.~Veps\"al\"ainen,
  {\it Dimensionally regularized Polyakov loop correlators in hot QCD,}
  JHEP {01} (2010) 054
  [0911.3480].
  %%CITATION = JHEPA,1001,054;%%

\bibitem{kgw}
  K.G.~Wilson and W.~Zimmermann,
  {\it Operator Product Expansions and Composite Field Operators in the General
  Framework of Quantum Field Theory,}
  Commun.\ Math.\ Phys.\  {24} (1972) 87.
  %%CITATION = CMPHA,24,87;%%

\bibitem{hbm_b}
  H.B.~Meyer,
  {\it Density, short-range order and the quark-gluon plasma,}
  Phys.\ Rev.\  D {79} (2009) 011502
  [0808.1950].
  %%CITATION = PHRVA,D79,011502;%%

\bibitem{sch}
  S.~Caron-Huot,
  {\it Asymptotics of thermal spectral functions,}
  Phys.\ Rev.\  D {79} (2009) 125009
  [0903.3958].
  %%CITATION = PHRVA,D79,125009;%%

\bibitem{hbm_c}
  N.~Iqbal and H.B.~Meyer,
  {\it Spatial correlators in strongly coupled plasmas,}
  JHEP {11} (2009) 029
  [0909.0582].
  %%CITATION = JHEPA,0911,029;%%

\bibitem{rhoE}
  Y.~Burnier, M.~Laine, J.~Langelage and L.~Mether,
  {\it Colour-electric spectral function at next-to-leading order,}
  JHEP {08} (2010) 094
  [1006.0867].
  %%CITATION = JHEPA,1008,094;%%

\bibitem{old}
  A.L.~Kataev, N.V.~Krasnikov and A.A.~Pivovarov,
  {\it Two-loop calculations for the propagators of gluonic currents,}
  Nucl.\ Phys.\  B {198} (1982) 508
  [Erratum-ibid.\  B {490} (1997) 505]
  [hep-ph/9612326].
  %%CITATION = NUPHA,B198,508;%%

\bibitem{ml}
  M.~L\"uscher,
  {\it Topological effects in QCD and the problem of short-distance
  singularities,}
  Phys.\ Lett.\  B {593} (2004) 296
  [hep-th/0404034].
  %%CITATION = PHLTA,B593,296;%%


\bibitem{jan}
  J.~M\"oller and Y.~Schr\"oder,
  {\it Open problems in hot QCD,}
  1007.1223.
  %%CITATION = ARXIV:1007.1223;%%

\bibitem{ope}
 V.A.~Novikov, M.A.~Shifman, A.I.~Vainshtein and V.I.~Zakharov,
 {\it Operator Expansion in Quantum Chromodynamics 
      Beyond Perturbation Theory,}
  Nucl.\ Phys.\  B {174} (1980) 378.
  %%CITATION = NUPHA,B174,378;%%

\bibitem{hbm_d}
  H.B.~Meyer,
  {\it The Bulk Channel in Thermal Gauge Theories,}
  JHEP {04} (2010) 099
  [1002.3343].
  %%CITATION = JHEPA,1004,099;%%

\bibitem{david}
  F.~David,
  {\it The Operator Product Expansion and Renormalons: A Comment,}
  Nucl.\ Phys.\  B {263} (1986) 637.
  %%CITATION = NUPHA,B263,637;%%


\bibitem{hbm_a}
  H.B.~Meyer,
  {\it Energy-momentum tensor correlators and spectral functions,}
  JHEP {08} (2008) 031
  [0806.3914].
  %%CITATION = JHEPA,0808,031;%%


\bibitem{hbm_e}
  H.B.~Meyer,
  {\it Lattice Gauge Theory Sum Rule for the Shear Channel,}
  Phys.\ Rev.\  D {82} (2010) 054504
  [1005.2686].
  %%CITATION = PHRVA,D82,054504;%%

\bibitem{rs}
  P.~Romatschke and D.T.~Son,
  {\it Spectral sum rules for the quark-gluon plasma,}
  Phys.\ Rev.\  D {80} (2009) 065021
  [0903.3946].
  %%CITATION = PHRVA,D80,065021;%%


\bibitem{ellis}
  P.J.~Ellis, J.I.~Kapusta and H.B.~Tang,
  {\it Low-energy theorems for gluodynamics at finite temperature,}
  {Phys.\ Lett.\  B} {443} (1998) 63
  [{nucl-th/9807071}].
  %%CITATION = PHLTA,B443,63;%%



\bibitem{hbm_f}
  H.B.~Meyer,
  {\it Finite Temperature Sum Rules in Lattice Gauge Theory,}
  Nucl.\ Phys.\  B {795} (2008) 230
  [0711.0738].
  %%CITATION = NUPHA,B795,230;%%



\bibitem{gsixg}
  K.~Kajantie, M.~Laine, K.~Rummukainen and Y.~Schr\"oder,
  {\it The pressure of hot QCD up to $g^6 \ln (1/g)$},
  Phys.\ Rev.\ D 67 (2003) 105008
  [hep-ph/0211321]. 
  %%CITATION = PHRVA,D67,105008;%%

\bibitem{es}
  %% 
  %% 1st paper but wrong answer:
  %%
  E.V.~Shuryak,
  {\it Theory of hadronic plasma,}
  Sov.\ Phys.\ JETP {47} (1978) 212.
  %%[Zh.\ Eksp.\ Teor.\ Fiz.\  {74} (1978) 408];
  %%CITATION = SPHJA,47,212;%%
  %%

\bibitem{chin}
  S.A.~Chin,
  {\it Transition to hot quark matter in relativistic heavy ion collision,}
  Phys.\ Lett.\ B {78} (1978) 552.
  %%CITATION = PHLTA,B78,552;%%









\end{thebibliography}
\end{document}